\documentclass[notitlepage,nofootinbib,preprintnumbers,amssymb,superscriptaddress,eqsecnum,a4paper,11pt]{article}
\usepackage{amsfonts,amssymb,amsmath,graphicx,color,bm,ulem,jheppub}
\definecolor{ultramarine}{rgb}{0.07, 0.04, 0.56}
\definecolor{cadmiumgreen}{rgb}{0.0, 0.42, 0.24}
\definecolor{indigo(dye)}{rgb}{0.0, 0.25, 0.42}
\usepackage[linktocpage=true]{hyperref}
\hypersetup{
colorlinks=true,
citecolor=ultramarine,
linkcolor=cadmiumgreen,
urlcolor=indigo(dye),
}

\newcommand{\f}[2]{\frac{#1}{#2}}  
\newcommand{\mk}[1]{\left( #1 \right)}

\newcommand{\be}{\begin{equation}}  
\newcommand{\ee}{\end{equation}}
\newcommand{\bem}{\begin{pmatrix}}
\newcommand{\eem}{\end{pmatrix}}
\newcommand{\rddots}{\reflectbox{$\ddots$}}

\newcommand{\1}{\boldsymbol{1}}
\newcommand{\A}{\mathcal{A}}
\newcommand{\N}{\mathcal{N}}
\newcommand{\I}{\mathcal{I}}

\newcommand{\E}{\mathcal{E}}
\newcommand{\mE}{\boldsymbol{\mathcal{E}}}

\newcommand{\mF}{\boldsymbol{F}}
\newcommand{\Q}{\boldsymbol{Q}}
\renewcommand{\P}{\boldsymbol{P}}
\renewcommand{\L}{\boldsymbol{\lambda}}
\newcommand{\R}{\boldsymbol{\rho}}
\newcommand{\bPhi}{\boldsymbol{\Phi}}
\newcommand{\bUp}{\boldsymbol{\Upsilon}}
\newcommand{\bOm}{\boldsymbol{\Omega}}
\newcommand{\0}{\boldsymbol{0}}
\newcommand{\pa}{\partial}
\newcommand{\eq}{{\rm eq}}

\newcommand{\vG}{{\boldsymbol G}}

\begin{document}

\title{  
Ghost-free theories with arbitrary higher-order time derivatives
}

\author[a,1]{Hayato Motohashi \note{Corresponding author.}}
\author[b]{Teruaki Suyama}
\author[b]{Masahide Yamaguchi}

\affiliation[a]{Center for Gravitational Physics, Yukawa Institute for Theoretical Physics, Kyoto University, Kyoto 606-8502, Japan}
\affiliation[b]{Department of Physics, Tokyo Institute of Technology,
2-12-1 Ookayama, Meguro-ku, Tokyo 152-8551, Japan}

\emailAdd{hayato.motohashi@yukawa.kyoto-u.ac.jp}
\emailAdd{suyama@phys.titech.ac.jp}
\emailAdd{gucci@phys.titech.ac.jp}

\abstract{
We construct no-ghost theories of analytic mechanics 
involving arbitrary higher-order derivatives in Lagrangian. 
It has been known that for theories involving at most second-order time derivatives in the Lagrangian,
eliminating linear dependence of canonical momenta in the Hamiltonian is 
necessary and sufficient condition to eliminate Ostrogradsky ghost.
In the previous work we showed for the specific quadratic model involving third-order derivatives 
that the condition is necessary but not sufficient, and linear dependence of canonical coordinates 
corresponding to higher time-derivatives also need to be removed appropriately. 
In this paper, we generalize the previous analysis and 
establish how to eliminate all the ghost degrees of freedom for general theories 
involving arbitrary higher-order derivatives in the Lagrangian.
We clarify a set of degeneracy conditions to eliminate all the ghost degrees of
freedom, under which we also show that the Euler-Lagrange equations are
reducible to a second-order system.
}

\preprint{YITP-18-32}

\maketitle


\section{Introduction}
\label{sec:intro}

The presence of inflation and the current accelerated expansion of the Universe
is strongly supported by observational results such as the cosmic
microwave background radiation anisotropies
\cite{Bennett:2012zja,Hinshaw:2012aka,Adam:2015rua,Ade:2015lrj} and type
Ia supernovae \cite{Riess:1998cb,Perlmutter:1998np}. One simple way to
explain these two regimes of accelerated expansion is to introduce additional degrees of freedom (DOFs) to
General Relativity and modify the law of gravitation.
In general, if one adds higher-than-first-order
derivative terms to an action, it leads to ghost DOFs
known as Ostrogradsky ghost~\cite{Ostrogradsky:1850fid}. Of course, even
if there is ghost DOF in a theory, it would not be
problematic as long as it appears above the scale one is interested in.
However, in cosmology one sometimes considers a
situation, in which higher derivative terms play dominant roles in the
dynamics. In such a case, the effective theory view point would be
invalidated and ghost DOFs must be removed to guarantee
healthiness and/or predictability of the theory. One of such famous
examples is Horndeski theory~\cite{Horndeski:1974wa} (equivalent to
generalized Galileon~\cite{Deffayet:2011gz,Kobayashi:2011nu}), which is
the most general single-field scalar-tensor theory whose Euler-Lagrange
equations of motion (EOMs) are up to second-order in derivatives, 
and thus free from additional ghost DOFs.

It was recognized recently that the requirement of second-order
Euler-Lagrange equations is too strong to avoid ghost
DOFs~\cite{Zumalacarregui:2013pma,Gleyzes:2014dya,Gleyzes:2014qga}.
This is because the highest orders of derivatives in the Euler-Lagrange equations
do not necessarily give a correct number of initial conditions:
Even if Euler-Lagrange equations directly derived from an action a priori include
higher-order time-derivative terms, 
there is no ghost instability as long as they can be recast into second-order system 
without introducing extra variables.
Ghost-free
condition can be thus investigated in a more explicit way in the
Hamiltonian picture. 
Under the assumption that a Lagrangian involves multiple variables and all the variables have up to $n$-th order derivatives ($n\geq 2$),
the Ostrogradsky theorem states that if the Lagrangian is nondegenerate with respect to the highest order derivatives, 
the Hamiltonian is unbounded due to the existence of ghost DOFs, which develops instabilities when the system couples to normal systems~\cite{Ostrogradsky:1850fid}.

One may then expect that the Ostrogradsky ghost can be removed by requiring the degeneracy of Lagrangian with respect to the highest-order derivatives, 
which corresponds to a removal of the highest $2n$-th order derivatives in the Euler-Lagrange equation.
However, evading the Ostrogradsky theorem is not sufficient to construct healthy models ({\it i.e.}\ no ghosts) as it is just a
statement of the sufficient condition for the existence of ghosts that non-degeneracy with respect to the highest-order derivatives inevitably
leads to ghosts.
In other words, degeneracy with respect to the highest-order derivatives does not guarantee the absence of all ghost DOFs.
In fact, it was demonstrated in~\cite{Motohashi:2014opa} that there exists a class of Lagrangians with up to $n$-th order derivatives that 
satisfies the degeneracy with respect to the highest-order derivative but 
ends up with
unbounded Hamiltonian 
due to the ghost DOFs associated with $(2n-1)$-th order derivatives in the Euler-Lagrange equation. 
Definitely, one needs more degeneracy conditions to eliminate all the ghost DOFs.
Another important point is that it is necessary for application to general theories of modified gravity to go beyond the assumption that all the variables have the same order of derivatives in Lagrangian, and to consider Lagrangian with several types of variables with different orders of derivatives.  
With several types of variables of different orders of derivatives in Lagrangian, degeneracy conditions are more nontrivial.

In \cite{Langlois:2015cwa}, the degeneracy condition was clarified for a specific Lagrangian for the quadratic-order model of degenerate higher-order scalar-tensor (DHOST) theories, which involves ``regular'' variables with at most first-order derivative and single ``special'' variable with at most second-order derivatives.
The degeneracy condition for general Lagrangian with multiple regular and special variables was developed in~\cite{Motohashi:2016ftl}.  
The degeneracy condition derived in~\cite{Motohashi:2016ftl} applies to any model involving up to second-order derivative terms in time.
Indeed, to the best of our knowledge, all of theories of modified gravity discussed so far include only up to second-order derivative
terms in time.  Specifically, Horndeski derived the most general second-order Euler-Lagrange equations for single-field scalar-tensor
theory, and then reconstructed the action starting from an action with
arbitrary finite order derivative terms, but the obtained action includes only up to second-order derivatives~\cite{Horndeski:1974wa}.
Gao proposed another extension, which involves arbitrary higher-order
derivatives in space but up to second-order derivatives in time~\cite{Gao:2014soa}.  
Theories beyond Horndeski~\cite{Gleyzes:2014dya,Gleyzes:2014qga} and degenerate higher-order scalar-tensor
theories~\cite{Langlois:2015cwa,BenAchour:2016fzp} also involve up to second-order derivatives.  
Thus, as far as we know, there is no explicit example in the context of field theory,
which includes third (or even higher) order derivatives in time but can avoid ghost instabilities.\footnote{It should be noted that some
of infinite-order derivative (non-local) theories can avoid ghost instabilities at least classically. For example, see
Refs.~\cite{Barnaby:2007ve,Barnaby:2008tc} and references therein.}

In the previous work~\cite{Motohashi:2017eya}, we provided a specific model which is quadratic in variables and 
involves third-order time derivatives in the Lagrangian. 
Our finding is that elimination of the canonical momenta in the Hamiltonian 
by the constraints and degeneracy conditions does not kill all the ghost DOFs associated with the higher derivatives and the ghost DOFs still remain.
Although the remaining ghost DOFs are hidden in a very nontrivial way in the Hamiltonian, in the case of the quadratic model,
canonical transformation makes those ghost DOFs manifest themselves as linear terms of canonical coordinates~\cite{Motohashi:2017eya}.
Presence of additional ghosts not in the form of the linear terms of the canonical momenta is a crucial difference from
theories involving at most second-order time derivatives.
We derived in~\cite{Motohashi:2017eya} a set of degeneracy conditions for the quadratic model, 
and confirmed that the Hamiltonian equations as well as the Euler-Lagrange equations are reducible 
to a system of second-order differential equations when the degeneracy conditions are imposed.

In this paper, we further generalize the previous analysis for theories 
involving at most second-order derivatives performed in~\cite{Motohashi:2016ftl} (see also \cite{Klein:2016aiq} for a similar analysis, \cite{Kimura:2017gcy} for a case including
fermionic degree of freedom, and
\cite{Langlois:2015cwa,Crisostomi:2016czh,Kimura:2016rzw,BenAchour:2016fzp,Crisostomi:2017aim,Crisostomi:2017ugk}
for field theoretical extensions),
as well as the previous analysis for the specific theory involving third-order derivatives in~\cite{Motohashi:2017eya}. 
Since the degeneracy conditions obtained in~\cite{Motohashi:2017eya} only apply to the specific quadratic model involving at most third-order derivatives, in the present paper,
we first clarify a set of degeneracy conditions for general Lagrangian involving third-order derivatives.
We also confirm that the Euler-Lagrange equations can be reduced into a second order differential equations.
Furthermore, we consider general Lagrangian involving arbitrary higher-order derivatives, 
and derive a set of degeneracy conditions, under which we confirm that
the Euler-Lagrange equations are reducible into second-order system.
Our result applies to any form of Lagrangian involving any higher-order derivatives. 
Thus, it is an important first step for construction of 
ghost-free theories of modified gravity with third- and even higher-order derivatives.

The organization of the rest of the paper is as follows.  
In \S\ref{sec:exs} we provide 
an explicit example
which includes arbitrary
higher-order derivatives in a Lagrangian but does not have Ostrogradsky ghosts. 
In \S\ref{sec:multi}, 
we investigate general Lagrangian involving three set of multiple variables with 
at most first-, second-, and third-order derivatives, respectively.
We derive a set of conditions to avoid Ostrogradsky ghosts. 
We show that with these conditions the Euler-Lagrange equations
are reducible to second-order system.  
Some of them are satisfied identically for single variable case, 
which is supplemented in the Appendix.
Finally in \S\ref{sec:multi-high} we extend our analysis to general Lagrangian
with arbitrary finite higher-order derivatives.  \S\ref{sec:conc} is
devoted to conclusions and discussion.

\section{Examples}
\label{sec:exs}

The specific example of ghost-free theory of quadratic model involving third-order derivatives is presented in~\cite{Motohashi:2017eya}.
In this section we provide an example of ghost-free theory involving arbitrary finite higher-order derivatives in Lagrangian.
We show that the Euler-Lagrange equations are rearrangeable to second-order system, 
and that through the Hamiltonian analysis the system does not possess the Ostrogradsky ghosts.

We consider the following Lagrangian   
\be \label{Ltoy} L = \f{1}{2} \f{\dot q^2}{1+\phi^{(d+1)}} + \f{1}{2}\dot\phi^2 , \ee
where $q=q(t)$, $\phi=\phi(t)$, and 
$\phi^{(d+1)}$ represents the $(d+1)$-th derivative of $\phi(t)$ with $d \ge 1$ being an integer.
This model is a generalization of the toy model considered in Sec.~7.1 of~\cite{Gabadadze:2012tr}, which corresponds to $d=1$ case.
The Euler-Lagrange equations for $q$ and $\phi$ are given by 
\begin{align}
\f{d}{dt} \mk{ \f{\dot q}{1+\phi^{(d+1)}} } &= 0 , \\
\ddot \phi + (-1)^{d+1} \f{1}{2} \f{d^{d+1}}{dt^{d+1}} \mk{ \f{\dot q^2}{(1+\phi^{(d+1)})^2 } } &= 0.  
\end{align}
Despite the appearance of higher derivative terms, we can see this system of equations is actually second-order system as follows.  
From the first equation we note that $\f{\dot q}{1+\phi^{(d+1)}} = {\rm const}$.
Plugging it to the second equation, we obtain $\ddot \phi=0$, with which the first equation reduces to $\ddot q=0$.  
Therefore the system is equivalent to   
\be \ddot q = \ddot \phi = 0, \ee 
which is clearly a second-order system for $2$ variables $q,\phi$ and requires $4$ initial conditions for $\{ q,\dot q, \phi,\dot \phi \}$. 
It is straightforward to consider a generalization of the model where the Lagrangian is given by a sum of \eqref{Ltoy} for multiple $q,\phi$ variables with different orders of derivatives.

Let us check the number of DOFs and the absence of Ostrogradsky ghost for the system \eqref{Ltoy} by Hamiltonian analysis.
By introducing auxiliary variables $Q^i$ and Lagrange multipliers $\lambda_i$, we rewrite the Lagrangian $L$ in \eqref{Ltoy} to an equivalent form
\begin{align} \label{Leqtoy}
L_{\rm eq} &= L(\dot q, \dot Q^d,Q^1) + \sum_{i=0}^{d-1} \lambda_i (\dot Q^i - Q^{i+1}) , \notag\\
L(\dot q, \dot Q^d,Q^1)&=\f{1}{2} \f{\dot q^2}{1 + \dot Q^d } + \f{1}{2}(Q^1)^2 ,
\end{align}
where we define $Q^0\equiv \phi$.
This Lagrangian yields at most second-order EOMs for $2(d+1)$ variables, $\{ q, Q^d, Q^i, \lambda_i \}$ with $i=0,\cdots,d-1$.  
Thus, a priori this system requires $4(d+1)$ initial conditions.
The form of $L_{\rm eq}$ allows us to define the canonical momenta for $\{ q, Q^d, Q^i, \lambda_i \}$ in the standard way:
\be
p = L_{\dot q} = \f{\dot q}{1 + \dot Q^d } , \quad
P_d = L_{\dot Q^d} = -\f{1}{2}\mk{ \f{\dot q}{1 + \dot Q^d } }^2 ,\quad
P_i = \lambda_i , \quad
\rho_i = 0 .
\ee
The last two equations are primary constraints associated with the introduction of auxiliary variables.
In addition to them, we note that there is an additional primary constraint $P_d=F(p)\equiv -p^2/2$.
In total, the primary constraints are
\be \Phi_i \equiv P_i - \lambda_i \approx 0, \quad 
\bar \Phi_i \equiv \rho_i \approx 0, \quad 
\Psi \equiv P_d - F \approx 0 . \ee
Time evolution of the canonical variables is governed by the total Hamiltonian, which is given by
\begin{align} 
H_T &= H + \mu_i \Phi_i + \bar \mu_i \bar \Phi_i + \nu \Psi ,\notag\\
H &= H_0 + \sum_{i=0}^{d-1} P_i Q^{i+1} \notag\\
H_0 &= \dot Q^d P_d + \dot q p - L,
\end{align}
where $\mu_i, \bar\mu_i, \nu$ are Lagrange multipliers.
The linear terms $\sum P_iQ^{i+1}$ correspond to the Ostrogradsky ghosts, by which the Hamiltonian is unbounded.

Since the primary constraints need to be satisfied through time evolution, we require time derivative of the primary constraints remain vanishing as consistency condition.
From the consistency condition $\dot{\bar\Phi}_i\approx 0$ and $\dot\Phi_i\approx 0$, we respectively obtain 
\be \mu_i \approx 0 , \quad \bar\mu_i \approx \{ \Phi_i , H \} + \nu \{ \Phi_i, \Psi \} . \ee
The consistency condition for $\Psi$ is given by
\be \label{conPsitoy} 0 \approx \dot\Psi = \{\Psi,H\} + \nu\{ \Psi, \Psi \} . \ee
Needless to say, the last term vanishes identically, but we kept it for later convenience.  Actually, the fact that this term identically vanishes means that this system satisfies the second degeneracy condition
[see \eqref{deg2I}].
From \eqref{conPsitoy} we obtain a secondary constraint
\be  - \{ \Psi, H \} = P_{d-1} \approx 0. \ee
We then check the consistency condition $0=\dot P_{d-1} = \{P_{d-1},H\} + \nu\{ P_{d-1}, \Psi \}$ and obtain a tertiary constraint $P_{d-2}=0$. Actually, it is clear from the linear terms $\sum P_iQ^{i+1}$ in the Hamiltonian that we successively obtain the constraints
\be P_{d-1} \approx 0,\quad 
P_{d-2} \approx 0, \quad ,
\cdots, \quad
P_1 \approx 0 . \ee
Finally the consistency condition for $P_1=0$ gives
\be 0 \approx \dot P_1 = P_0 - Q_1 , \ee
which is the last constraint as its consistency condition is identically satisfied.
Clearly, the constraints remove linear terms in the Hamiltonian, and thus eliminate the Ostrogradsky ghosts.

Hence we expect the system possesses only healthy $2$ DOFs.
To count the number of DOFs, we classify all the constraints
obtained above to first class and second class by checking the Poisson
brackets between them, which form the Dirac matrix.  The Dirac matrix is
given by \be
\begin{array}{l|ccccccc}
  & \Phi_j & \bar\Phi_j & \Psi & P_{d-1} & \cdots & P_1 & P_0-Q_1 \\ \hline
  \Phi_i     & 0 & -\1 &  \\
  \bar\Phi_i & \1 & 0 &  \\
  \Psi       &  \\
  P_{d-1}    &  \\
  \vdots     &  \\
  P_1        &  &  &  &  &  & 0 & 1 \\
  P_0-Q_1    &  &  &  &  &  & -1 & 0
\end{array} 
\ee
where $\1$ is the unit matrix and blank arguments are zeros.
Hence we end up with $2d+2$ second class constraints $\Phi_i,\bar\Phi_i,P_1, P_0-Q_1$, and $d-1$ first class constraints $\Psi, P_{d-1}, P_{d-2}, \cdots, P_2$.  
Starting the primary first class constraint $\Psi$, we can check the Dirac test is satisfied:  
Since the chain of the Poisson brackets exhausts all first class constraints as $\{ H,\Psi \} = P_{d-1}, \{ H,P_{d-1} \} = P_{d-2}, \cdots , \{ H,P_3 \} = P_2$, all the first class constraints are generator of gauge transformations. 
Therefore, the number of DOFs for the system is given by $[4(d+1) - (2d+2) - 2(d-1)]/2=2$, which is consistent with the Euler-Lagrange picture.

\section{Lagrangian with multiple third-order derivatives}
\label{sec:multi}

The example in \S\ref{sec:exs} shows that it is indeed possible to involve arbitrary higher-order derivatives in Lagrangian and construct no-ghost theory.
In this case, some part of degeneracy conditions could be identically satisfied 
due to the particular form of the Lagrangian.
For more general Lagrangians, we need to impose a certain set of degeneracy conditions, for which it is worthwhile to remind the lesson obtained in \cite{Motohashi:2017eya}.
In \cite{Motohashi:2017eya}, we investigated the quadratic model involving third-order derivatives and clarified that it is necessary to impose a sufficient number of degeneracy conditions to eliminate all ghost DOFs. 
In particular, fixing linear terms in conjugate momenta in the Hamiltonian is not sufficient as linear terms in canonical coordinates themselves lurk in the Hamiltonian in a nontrivial way. 
We need to impose degeneracy conditions and continue the Dirac algorithm until we are left with healthy DOFs whose number matches that of variables.
The final goal of the present paper is to generalize this process for general Lagrangian involving arbitrary higher-order derivatives (see \S\ref{sec:multi-high}).

In this section, we consider Lagrangian involving multiple
variables $\psi^n(t)$ with third-order derivatives
and multiple regular variables $q^i(t)$:
\be \label{LagI} L(\dddot\psi^n, \ddot\psi^n, \dot\psi^n, \psi^n;
\ddot\phi^a, \dot\phi^a, \phi^a; \dot q^i, q^i), \ee
where $n,a,i$ run from $1$ to $\N, \A, \I$, respectively.
In order to cover a wide class of Lagrangians up to the third-order time derivatives,
we also include the variables $\phi^a$ that enter the Lagrangian up to their second-order time derivatives.
We investigate the Hamiltonian analysis in \S\ref{ssec:Ham2} to derive degeneracy conditions, and the Euler-Lagrange equations in \S\ref{ssec:ELeq2} to show the reduction to second-order system.
For the special case $\N=1$ and $\A=0$, some part of degeneracy conditions are identically satisfied, for which we provide a brief explanation in Appendix~\ref{sec:sing}. 
Instead of dealing with the Lagrangian \eqref{LagI}, for the practical purpose, we consider an equivalent Lagrangian given by
\begin{align} \label{LageqI} 
L_\eq &\equiv L(\dot Q^n, Q^n, R^n, \psi^n; \dot Q^{\N+a}, Q^{\N+a}, \phi^a; \dot q^i, q^i) \notag\\
&~~~+ \xi_n (\dot\psi^n - R^n) + \lambda_n (\dot R^n - Q^n) + \lambda_{\N+a}
(\dot \phi^a - Q^{\N+a}) , 
\end{align}
and denote $Q^I = (Q^n, Q^{\N+a} )$.

\subsection{Hamiltonian analysis}
\label{ssec:Ham2}

The canonical momenta for $Q^I,q^i,R^n,\psi^n,\phi^a,\xi_n,\lambda_I$ are respectively given by
\be 
P_{Q^I}= L_I, \quad
p_i = L_i, \quad
P_{R^n} = \lambda_n, \quad
\pi_{\psi^n} = \xi_n, \quad
\pi_{\phi^a} = \lambda_{\N+a}, \quad
\rho_{\xi_n}=0, \quad
\rho_{\lambda_I}=0, \label{3-multi-constra}\ee
where $L_I\equiv \pa L/\pa \dot Q^I$ and $L_i\equiv \pa L/\pa \dot q^i$.
Below we simply write $P_{Q^I}\to P_I$ when we denote all $I=(n,a)$ components, whereas we retain the notation $P_{Q^n}$ for $n$ components to distinguish it from $P_{R^n}$.
The number of canonical variables are a priori $10\N+6\A+2\I$.

From the latter six equations, we obtain $4\N+2\A$ primary constraints
\begin{align}
&\Phi_n\equiv P_{R^n}-\lambda_n \approx 0, \quad
\Phi_{\N+a}\equiv \pi_{\phi^a} - \lambda_{\N+a} \approx 0, \quad 
\Phi_{\N+\A+n}\equiv \pi_{\psi^n} - \xi_n \approx 0, \notag\\
&\bar\Phi_{n}\equiv \rho_{\lambda_n} \approx 0, \quad\quad\quad~~
\bar\Phi_{\N+a}\equiv \rho_{\lambda_{\N+a}} \approx 0, \quad\quad\quad~~
\bar\Phi_{\N+\A+n}\equiv \rho_{\xi_n} \approx 0.
\end{align}
At this moment, it is nontrivial whether the first two equations in \eqref{3-multi-constra} provide further constraints or not. 
However, if they do not provide constraints, the system has DOF more than the number of variables, and we end up with Ostrogradsky ghost.
We thus assume the existence of an additional primary constraint in the following way. 
Let us consider the infinitesimal changes of $P_I, p_i$, which are related as 
\be \label{dPI}
\bem \delta P_I - L_{I x} \delta x \\ \delta p_i - L_{i x} \delta x \eem =
K
\bem \delta \dot Q^J \\ \delta \dot q^j \eem ,
\ee
where the kinetic matrix $K$ is given by  
\be K\equiv \bem 
L_{IJ} & L_{Ij} \\
L_{iJ} & L_{ij} 
\eem , \ee
and $x = (Q^I,R^n,\psi^n,\phi^a,q^i)$, and summation for overlapping $x$ is implicit.
If $\det K\neq 0$, one can locally express $\dot Q^I, \dot q^i$ in terms of canonical variables, meaning that there is no further primary constraint.  
Therefore, we require $\det K=0$.
More precisely, we require the maximal degeneracy of 
the part of $K$ corresponding to the higher derivatives to eliminate ghost DOFs.
On the other hand,
to avoid eliminating DOFs coming from $q^i$, we assume 
\be \det k \neq 0 , \ee
where $k_{ij}$ is a sub-kinetic matrix defined by $k_{ij} \equiv L_{ij}$.
Under this assumption, $K$ can be rewritten as 
\be \label{KreI} K = R
\bem 
L_{IJ} - L_{Ii} k^{ij} L_{jJ} & 0 \\
0 & k 
\eem
S,
\ee
where $k^{ij}$ is the inverse matrix of $k_{ij}$ and
\be R \equiv \bem
1 & A^T \\
0 & 1 
\eem, \quad 
S \equiv \bem
1 & 0 \\ 
A & 1 
\eem,\quad 
{A^i}_I\equiv k^{ij}L_{jI}.
\ee
Now it is clear that the maximal degeneracy of the part of $K$ corresponding to the higher derivatives implies
\be \label{deg1I} L_{IJ} - L_{Ii} L^{ij} L_{jJ} = 0 , \ee
which is the first degeneracy condition we impose.
Under this condition, 
\eqref{KreI} reads
(see also Appendix~B.3 of \cite{Motohashi:2016ftl})
\be \label{KdiagI} K = R \bem 
0 & 0 \\
0 & k 
\eem S . \ee

Plugging \eqref{KdiagI} to \eqref{dPI} we obtain
\begin{align} \label{ddQddqI}
\delta P_I - L_{Ii} L^{ij} \delta p_j &= ( L_{Ix} - L_{Ii} L^{ij} L_{jx}) \delta x ,\notag\\
L^{ij}L_{Ij} \delta \dot Q^I + \delta \dot q^i &= L^{ij} (\delta p_j - L_{jx} \delta x) .
\end{align}
We thus obtain additional primary constraints
\be \label{PsiI} \Psi_I \equiv P_I - F_I (p_i, x) \approx 0, \ee
with
\be \label{FIrel} F_{Ip_i} = L^{ij} L_{Ij} ,\quad
F_{Ix} = L_{Ix} - F_{Ip_i} L_{ix} . \ee

The total Hamiltonian is given by 
\begin{align} \label{HamI}
H_T &= H + \mu_\alpha \Phi_\alpha + \bar\mu_{\alpha} \bar\Phi_{\alpha} + \nu_I \Psi_I ,\notag\\
H &= H_0 + P_{R^n} Q^n + \pi_{\psi^n} R^n + \pi_{\phi^a} Q^{\N+a} ,\notag\\
H_0 &= \dot Q^I P_I + \dot q^i p_i - L(\dot Q^n, Q^n, R^n, \psi; \dot Q^{\N+a}, Q^{\N+a}, \phi^a; \dot q^i, q^i),
\end{align}
where $\Phi_\alpha=(\Phi_n,\Phi_{\N+a},\Phi_{\N+\A+n})$ with $\alpha=1,\cdots,2\N+\A$ 
and so does $\bar\Phi_\alpha$, and $\mu_\alpha,\bar\mu_\alpha,\nu_I$ are the Lagrange multipliers associated with the primary constraints $\Phi_\alpha, \bar\Phi_\alpha,\Psi_I$, respectively.
The momenta $P_{R^n}, \pi_{\psi^n}, \pi_{\phi^a}$ show up in the Hamiltonian only through the linear terms, which lead to the Ostrogradsky instability.
We shall see that the secondary constraints fix $P_{R^n}, \pi_{\phi^a}$, and the tertiary constraints fix $\pi_{\psi^n}$.

To guarantee that the primary constraints $\Phi_\alpha, \bar\Phi_\alpha,\Psi_I$ 
are satisfied through time evolution, 
the consistency conditions $\dot \Phi_\alpha\approx 0, \dot{\bar\Phi}_\alpha\approx 0, \dot \Psi_I\approx 0$ should be satisfied.
From $\dot\Phi_\alpha \approx 0$, we obtain equations for $\bar\mu_\alpha$ as 
\be \bar\mu_\alpha \approx \{ \Phi_\alpha, H \} + \nu_I \{ \Phi_\alpha, \Psi_I \} , \ee
which read
\begin{align} \label{mueq1I} 
\bar\mu_n &\approx -\pi_{\psi^n} + L_{R^n} + \nu_I \{ \Phi_n, \Psi_I \} ,\notag\\
\bar\mu_{\N+a} &\approx L_{\phi^a} + \nu_I \{ \Phi_{\N+a}, \Psi_I \} ,\notag\\
\bar\mu_{\N+\A+n} &\approx L_{\psi^n} + \nu_I \{ \Phi_{\N+\A+n}, \Psi_I \} .
\end{align}
On the other hand, 
$\dot{\bar\Phi}_{\alpha}\approx 0$ fixes $\mu_\alpha$ as
\be \label{mueq2I} \mu_\alpha \approx 0  .\ee  
Therefore the consistency conditions for the primary constraints $\Phi_\alpha,\bar\Phi_\alpha$ 
determine Lagrange multipliers $\bar\mu_\alpha,\mu_\alpha$, respectively, 
and do not generate secondary constraints.
The remaining consistency conditions for the primary constraints $\Psi_I$ are
\be \label{conPsiI} 0 \approx \dot\Psi_I = \{ \Psi_I, H \} + \nu_J \{ \Psi_I, \Psi_J \} ,  \ee
where we substituted \eqref{mueq2I}.
As shown in \cite{Motohashi:2014opa}, the appearance of the matrix $\{ \Psi_I, \Psi_J \}$ is the nature of the multi-variable system, and if $\{ \Psi_I, \Psi_J \}$ is nondegenerate, this system suffers from ghost DOFs. 
We thus need further constraints to eliminate them.
To make all the equations give secondary constraints, we impose the second degeneracy conditions
\be \label{deg2I} \{ \Psi_I, \Psi_J \} 
= F_{JQ^I} - F_{IQ^J} + F_{Iq^i} F_{Jp_i} - F_{Ip_i} F_{Jq^i} 
= 0 . \ee
Under the second degeneracy conditions~\eqref{deg2I} we obtain secondary constraints
\be \Upsilon_I \equiv -\{ \Psi_I, H\} \approx 0, \ee
which read
\begin{align} \label{con2I}
\Upsilon_n &= P_{R^n} - L_{Q^n} + F_{np_i}L_{q^i} + \dot x F_{nx} \equiv P_{R^n} - G_{n}, \notag\\ 
\Upsilon_{\N+a} &= \pi_{\phi^a} - L_{Q^{\N+a}} + F_{\N+a, p_i}L_{q^i} + \dot x F_{\N+a, x} \equiv \pi_{\phi^a} - G_{\N+a},
\end{align}
which fix $P_{R^n},\pi_{\phi^a}$, eliminating Ostrogradsky instability coming from terms linear in them in the Hamiltonian \eqref{HamI}.

Note that, for the case $\N=1$ and $\A=0$, the Poisson bracket is $\{ \Psi_I, \Psi_J \}\to \{\Psi,\Psi\}$ which identically vanishes.  Hence, as mentioned earlier, the degeneracy conditions corresponding to \eqref{deg2I} are identically satisfied, and one obtains the secondary constraints corresponding to \eqref{con2I} automatically.

We can show that $G_I = G_I(p_i,x)$ as follows.  
By using the second equation of \eqref{ddQddqI}, we can show that $\delta \dot Q^I$ and $\delta \dot q^i$ terms of the variation of $L_x - X_{Ip_i}L_{q^i} + \dot y X_{Iy}$ for general $X_I=X_I(p_i,x)$ can be given by  
\be \label{dXL2} 
\delta (L_x- X_{Ip_i}L_{q^i} - \dot y X_{Iy}) \supset (F_{Jx} - X_{IQ^J} + X_{Iq^i} F_{Jp_i} - X_{Ip_i} F_{Jq^i})\delta \dot Q^J  .  
\ee
Applying this relation to $G_I = L_{Q^I} - F_{Ip_i}L_{q^i} - \dot x F_{Ix}$, we obtain 
\be \label{GpI}
\delta G_I \supset ( F_{JQ^I} - F_{IQ^J} + F_{Jp_i}F_{Iq^i} - F_{Ip_i}F_{Jq^i} ) \delta \dot Q^J .
\ee
We see that the coefficient precisely coincides with the second degeneracy conditions \eqref{deg2I}.
We thus conclude $G_I=G_I(p_i,x)$.
For the case $\N=1, \A=0$, one can show the right hand side of \eqref{GpI} identically vanishes.

The consistency conditions for the secondary constraints are given by 
\be \label{crel2I} 0 \approx \dot \Upsilon_I = \{ \Upsilon_I, H \} + \nu_J \{ \Upsilon_I, \Psi_J \} . \ee
As mentioned earlier, among $\Upsilon_I = (\Upsilon_n, \Upsilon_{\N+a})$, the latter part are constraints eliminating Ostrogradsky ghost associated with $\pi_{\phi^a}$.
We thus would like to stop the reduction of $\phi$ sector, while we still need further constraints to eliminate Ostrogradsky ghost in $\psi$ sector.
Hence, we require $\det \{ \Upsilon_{\N+a}, \Psi_{\N+b} \} \neq 0$ by which $\nu_{\N+a}$ are fixed.
To remove ghost DOFs from $\psi$ sector under the condition $\det \{ \Upsilon_{\N+a}, \Psi_{\N+b} \} \neq 0$, 
one may be tempted to impose the third degeneracy conditions as
\be  \{ \Upsilon_n, \Psi_m \}- \{ \Upsilon_n, \Psi_{\N+a} \}
{ \{ \Upsilon_{\N+b},\Psi_{\N+a} \} }^{-1}
\{ \Upsilon_{\N+b}, \Psi_m \} =0,  \ee
so that $\{ \Upsilon_I, \Psi_J \}$ can be decomposed as
\be \{ \Upsilon_I, \Psi_J \} = R' \bem 
0 & 0 \\
0 & \{ \Upsilon_{\N+a}, \Psi_{\N+b} \}
\eem S' , \ee
with some nontrivial $R', S'$, in parallel to \eqref{deg1I} and \eqref{KdiagI}.
This time, for simplicity, we impose 
\be \label{deg3I}
\{ \Upsilon_n,  \Psi_I \}=F_{I R^n}-G_{nQ^I}+G_{nq^i} F_{Ip_i}-G_{np_i}F_{Iq^i}=0, \ee
as the third degeneracy conditions
to ensure the structure 
\be \label{UIPJ} \{ \Upsilon_I, \Psi_J \} = \bem 
0 & 0 \\
\{ \Upsilon_{\N+a}, \Psi_m \} & \{ \Upsilon_{\N+a}, \Psi_{\N+b} \}
\eem , \ee
where $\det\{ \Upsilon_{\N+a}, \Psi_{\N+b} \}\neq 0$.
Plugging \eqref{UIPJ} into \eqref{crel2I}, the first row yields the tertiary constraints given by
\be \label{con3I} 
0 \approx \Lambda_n \equiv - \{ \Upsilon_n, H \} = 
\pi_{\psi^n}-L_{R^n}+G_{np_i} L_{q^i}+{\dot x} G_{nx} \equiv \pi_{\psi^n}-I_n (p_i,x), \ee
where we have again used \eqref{dXL2} to show that $I_n = I_n(p_i,x)$. 
Thus, the tertiary constraints fix $\pi_{\psi^n}$.
On the other hand, the remaining $\A$ components of \eqref{crel2I} give $\A$ equations for 
\be \label{nueq1I} 
\{ \Upsilon_{\N+a}, \Psi_J \} \nu_J \approx -L_{\phi^a}+G_{\N+a,p_i} L_{q^i}+{\dot x} G_{\N+a,x}.
\ee
Since $\det \{ \Upsilon_{\N+a}, \Psi_{\N+b} \} \neq 0$, this equation fix $\nu_{\N+a}$ as expected.
We shall see in \eqref{EOMa} that the right hand side is vanishing by virtue of EOM for $\phi^a$.
For the case $\N=1, \A=0$, the Poisson bracket is $\{ \Upsilon_I, \Psi_J \}\to \{\Upsilon,\Psi\}$ and one simply needs to impose $\{\Upsilon,\Psi\}=0$ as the degeneracy condition.

Therefore, we have fixed all the linear momentum terms $P_{R^n}, \pi_{\psi^n}, \pi_{\phi^a}$ in the Hamiltonian \eqref{HamI}.
However, as demonstrated in \cite{Motohashi:2017eya} for the quadratic model, 
the salient feature that the Ostrogradsky ghosts are not completely eliminated
even after all the linear terms in momenta have been removed by the constraints is expected to be generic in the higher
derivative theories with more than second time-derivatives in the Lagrangian. 
This is because the canonical variables $Q^n$ correspond
to the second time derivatives of $\psi^n$ and could become the source of the Ostrogradsky ghosts.

In the present case with general Lagrangian, an explicit redefinition of variables that reveals the hidden ghost is not trivial. 
Instead, we use the counting of the number of phase space variables.
All the phase space variables of the current system~\eqref{LageqI} are
\be \label{phaseI}
\begin{array}{ccccccc}
  Q^I & q^i & R^n & \psi^n & \phi^a & \boxed{\xi_n} & \boxed{\lambda_I}   \\ 
  \boxed{ P_{Q^I } } & p_i & \boxed{P_{R^n}} & \boxed{\pi_{\psi^n}} & \boxed{\pi_{\phi^a}} & \boxed{\rho_{\xi_n}} & \boxed{\rho_{\lambda_I}}
\end{array} ,
\ee
where the boxed variables are fixed in terms of other variables via constraints obtained so far.
Therefore, we currently have $3\N+2\A+2\I$ free variables in phase space. 
The original Lagrangian~\eqref{LagI} depends on $\psi^n, \phi^a, q^i$ and 
we would like to have a theory such that these variables behave as if they are ``ordinary'' variables 
corresponding to $2(\N+\A+\I)$ free variables in phase space.  
Therefore, from \eqref{phaseI} the current system has $\N$ extra phase space variables, 
and we assume that they are the hidden Ostrogradsky ghosts, 
which do not appear in the Hamiltonian as linear momentum terms.
Generalizing the result obtained in \cite{Motohashi:2017eya}, 
we expect that for some simple cases it is possible to find out an explicit redefinition of variables to
reveal the hidden ghost as a term linear in $Q_n$ in the Hamiltonian.

Based on these considerations, to eliminate the hidden Ostrogradsky ghosts, 
we require that the consistency conditions for 
the tertiary constraints~\eqref{con3I} 
\begin{align} \label{consisLam} 
0 \approx \dot \Lambda_n &= \{ \Lambda_n, H_T \}=\{ \Lambda_n, H\}+\nu_J \{ \Lambda_n,\Psi_J \},
\end{align}
does not determine any Lagrange multipliers, and hence generate the quaternary constraints.
Along the same line as the third degeneracy condition~\eqref{deg3I}, 
as the simplest case, although not the most general,
we require
\be \label{deg4I}
\{ \Lambda_n,\Psi_I \}=F_{I \psi^n}-I_{nQ^I}+I_{nq^i} F_{Ip_i}-I_{np_i} F_{Iq^i}= 0, \ee
as the fourth degeneracy conditions.
Then, the consistency conditions~\eqref{consisLam} for $\Lambda_n$ yield the following quaternary constraints,
\be \label{con4I} 
0 \approx \Omega_n \equiv - \{ \Lambda_n, H \}=-L_{\psi^n}+I_{np_i}L_{q^i}+{\dot x}I_{nx},
\ee
which fix the $\N$ phase space variables, precisely matching the number of $Q_n$, as expected.
Again, using \eqref{dXL2} we can show that $\Omega_n = -J_n(p_i,x)$.

For the case $\N=1, \A=0$, one can show $\{ \Lambda, \Psi \}=0$ identically holds
(see Appendix~\ref{sec:sing} for the proof),
and the quaternary constraint is automatically obtained. 
This makes sense since the absence of such constraint would lead to the equations of motion 
containing third time-derivative of a single variable only, which is incompatible with the nature
of Euler-Lagrange equations.

The consistency conditions for $\Omega_n$ yield 
\be \label{nueq2I} \{ \Omega_n, \Psi_I \} \nu_I \approx J_{np_i}L_{q^i} + \dot x J_{nx}, \ee
whose right hand side shall be shown to be vanishing in \eqref{EOMnd} by virtue of time derivative of EOM for $\psi^n$.
Thus, \eqref{nueq1I} and \eqref{nueq2I} form a system of $\N+\A$ equations for $\nu_I$.
Since we have reduced the number of the unconstrained canonical variables to $2(\N+\A+\I)$, 
we do not impose further constraints.
In other words, we require all the Lagrange multipliers $\nu_I$ are determined by \eqref{nueq1I} and \eqref{nueq2I}. 
Denoting 
\be \hat\Omega_I\equiv (\Omega_n, \Upsilon_{\N+a}), \ee
and
\be Z_{IJ} \equiv \{ \hat\Omega_I , \Psi_J \}, \ee
we require each submatrix of $Z_{IJ}$ is nondegenerate: 
\be \label{lastmulti} \det Z_{ab} \neq 0 ,\quad \det Z_{nm} \neq 0.  \ee
Under this condition we obtain
\be \nu_I \approx 0 . \ee

The number of constraints is
\begin{align}
\Phi_\alpha ,\bar\Phi_\alpha &: 4\N+2\A , \notag\\
\Psi_I &: \N+\A ,\notag\\
\hat\Omega_I &: \N+\A , \notag\\
\Upsilon_n &: \N ,\notag\\
\Lambda_n &: \N ,
\end{align}
and the total number is thus $8\N+4\A$.
Using the definition $\Upsilon_I=-\{\Psi_I,H\}$ and the Jacobi identity we can show 
\begin{align} 
\{ \Upsilon_I, \Upsilon_n \} &= \{ \Lambda_n, \Psi_I \} + \{ \{ \Upsilon_n,\Psi_I\},H \} =0 , \notag\\ 
\{ \Upsilon_I, \Lambda_n \} &= \{ \Omega_n, \Psi_I \} + \{ \{ \Lambda_n,\Psi_I\},H \} = Z_{nI}, \notag\\
Z_{a n}&=\{ \Upsilon_{\N+a}, \Psi_n \}=\{ \Upsilon_n, \Psi_{\N+a} \}=0.
\end{align}
With this in mind, the Dirac matrix is given by
\be
\begin{array}{l|cccccc}
  & \Phi_\beta & \bar\Phi_\beta & \Psi_J & \hat\Omega_J & \Upsilon_m & \Lambda_m  \\ \hline
  \Phi_\alpha     & 0 & -\1 & * & * & * & * \\
  \bar\Phi_\alpha & \1 & 0 & 0 & 0 & 0 & 0  \\
  \Psi_I          & * & 0 & 0 & -Z_{JI} & 0 & 0   \\
  \hat\Omega_I    & * & 0 & Z_{IJ} & * & * & *  \\
  \Upsilon_n      & * & 0 & 0 & * & 0 & Z_{mn}  \\
  \Lambda_n       & * & 0 & 0 & * & -Z_{nm} & *  \\
\end{array} 
\ee
and the determinant of the Dirac matrix is given by
\be ( \det Z_{ab} )^2 (\det Z_{nm})^4 , \ee
which does not vanish by virtue of \eqref{lastmulti}. 
Therefore, since all the $8\N+4\A$ constraints are second class, the number of DOF is 
\be \f{1}{2} [ 10\N+6\A+2\I - (8\N+4\A) ] = \N+\A+\I . \ee

\subsection{Euler-Lagrange equation}
\label{ssec:ELeq2}

The Euler-Lagrange equation for \eqref{LageqI} is given by
\begin{align}
\label{ELeq1I} \dot L_i - L_{q^i} &= 0 ,\\
\label{ELeq2I} \dot L_I - L_{Q^I} + \lambda_I &= 0 ,\\
\label{ELeq3I} L_{R^n} - \xi_n - \dot\lambda_n &= 0, \\
\label{ELeq4I} L_{\psi^n} - \dot\xi_n &= 0 ,\\
\label{ELeq5I} L_{\phi^a} - \dot\lambda_{N+a} &= 0,\\
\label{ELeq6I} Q^n - \dot R^n &= 0,\\
\label{ELeq7I} R^n - \dot\psi^n &= 0,\\
\label{ELeq8I} Q^{N+a} - \dot \phi^a &= 0 .
\end{align}
To obtain EOM for $\psi^n,\phi^a,q^i$ we successively take time derivative of the Lagrange multipliers $\lambda_I,\xi_n$.

First, we begin with $\lambda_I$.
From \eqref{ELeq2I}, a priori $\lambda_I$ depends on $\ddot Q^I$ which we would like to avoid.
Using the first degeneracy condition \eqref{deg1I} or the additional primary constraints~\eqref{PsiI}, $L_I=F_I(L_i,x)$ with the relations \eqref{FIrel}, we can show \eqref{ELeq1I} and \eqref{ELeq2I} can be transformed as
\begin{align}
\label{lamI} &~~~~ \lambda_I = L_{Q^I} - F_{Ip_i} L_{q^i} - \dot x F_{Ix} , \\
\label{EOMi} \E_i &\equiv \ddot q^i + F_{Ip_i} \ddot Q^I - L^{ij} (L_{q^j} - \dot x L_{jx}) = 0 .
\end{align}
The first equation \eqref{lamI} corresponds to the secondary constraints \eqref{con2I}.

Second, we take time derivative of \eqref{lamI} to obtain $\xi_n$ from \eqref{ELeq3I}, and EOM for $\phi^a$ from \eqref{ELeq5I}.
Again, to avoid for them to depend on $\ddot Q^I$, we impose $\lambda_I=G_I(L_i,x)$.
Indeed, in \eqref{GpI} we showed it holds by virtue of the second degeneracy condition \eqref{deg2I}.
Thus $\xi_n$ and EOM for $\phi^a$ does not depend on $\ddot Q^I$. 
In fact, from \eqref{ELeq3I} we obtain 
\be \label{xiI} \xi_n = L_{R^n} - G_{np_i} L_{q^i} - \dot x G_{nx} \ee
which corresponds to the tertiary constraints \eqref{con3I}.
Also, from \eqref{ELeq5I} we obtain EOM for $\phi^a$
\be \label{EOMa} \E_a \equiv L_{\phi^a} - G_{\N+a,p_i} L_{q^i} - \dot x G_{\N+a,x} = 0 , \ee
which corresponds to the right hand side of \eqref{nueq1I}.

Third, we take time derivative of \eqref{xiI} to obtain EOM for $\psi^n$ from \eqref{ELeq4I}. 
Again, to avoid its $\ddot Q^I$ dependency, we impose $\xi_n=I_n(L_i,x)$, which has been actually shown in the previous subsection
by using the third degeneracy condition \eqref{deg3I}.
From \eqref{ELeq4I} we obtain EOM for $\psi^n$
\be \label{EOMn} \E_n \equiv L_{\psi^n} - I_{np_i} L_{q^i} - \dot x I_{nx} = 0 , \ee
which corresponds to the quaternary constraints \eqref{con4I}.

We thus obtain EOM for $q^i,\phi^a,\psi^n$ as \eqref{EOMi}, \eqref{EOMa}, \eqref{EOMn}, but they still contain higher derivatives.
Below we construct a set of EOMs with derivatives up to second-order.

We derive another independent EOM by taking time derivative of \eqref{EOMn}.
To avoid its $\ddot Q^I$ dependency, we impose $\E_n=J_n(L_i,x)$ which holds by virtue of the fourth degeneracy condition~\eqref{deg4I}. 
Therefore,
\be \label{EOMnd} 0 = \dot J_n = J_{np_i} L_{q^i} + \dot x J_{nx} , \ee
which coincides with the right hand side of \eqref{nueq2I}.
Generalizing the derivation of Eq.~(24) from Eq.~(23) in \cite{Motohashi:2017eya}, we expect that in general the condition \eqref{lastmulti} guarantees that we can solve
\eqref{EOMa}--\eqref{EOMnd} for $\dot Q^n,Q^n,\dot Q_{\N+a}$ and obtain 
\begin{align} 
\label{dQnsol} \dot Q^n &= \dot Q^n (\dot q^i, Q^{\N+a}, R^n, \psi^n, \phi^a, q^i) , \\
\label{Qnsol} Q^n &= Q^n (\dot q^i, Q^{\N+a}, R^n, \psi^n, \phi^a, q^i) , \\
\label{dQasol} \dot Q^{\N+a} &= \dot Q^{\N+a} (\dot q^i, Q^{\N+a}, R^n, \psi^n, \phi^a, q^i) .
\end{align}
The equations \eqref{Qnsol}, \eqref{dQasol} are EOMs containing at most $\ddot\psi^n=Q^n$, $\ddot\phi^a=\dot Q^{\N+a}$, respectively. 
Taking time derivative of \eqref{dQnsol} and \eqref{dQasol}, and using these equations we obtain
\be \label{ddQIsol} \ddot Q^I = \ddot Q^{I} (\ddot q^i, \dot q^i, Q^{\N+a}, R^n, \psi^n, \phi^a, q^i) .  \ee
By substituting \eqref{dQnsol}--\eqref{ddQIsol} to \eqref{EOMi}, we obtain EOM containing at most $\ddot q^i$. 
Combining it with \eqref{Qnsol}, \eqref{dQasol}, we thus obtain a system of $\N+\A+\I$ EOMs that contain at most $\ddot\psi^n,\ddot\phi^a, \ddot q^i$.

\section{Lagrangian with arbitrary higher-order derivatives}
\label{sec:multi-high}

Finally we extend the analyses in
\S\ref{sec:multi} for the Lagrangian with
third-order derivatives to that with arbitrary higher order derivatives.
We explore the following Lagrangian involving arbitrary
higher $(d+1)$-th order derivatives:
\be L = L ( 
\phi^{i_0}, \dot\phi^{i_0}; 
\phi^{i_1}, \dot\phi^{i_1}, \ddot\phi^{i_1}; 
\phi^{i_2}, \dot\phi^{i_2}, \ddot\phi^{i_2}, \dddot\phi^{i_2}; 
\cdots ;
\phi^{i_d}, \dot\phi^{i_d}, \cdots ,\phi^{i_d(d+1) } ) . 
\ee
Here, the index $i_k$ counts the number of $\phi(t)$ variables and runs 
\begin{align}
i_0 &= 1, \cdots, n_0 ,\notag\\
i_1 &= n_{0}+1, \cdots , n_0 + n_1 ,\notag\\
&\vdots \notag\\
i_d &= \sum_{k=0}^{d-1} n_k + 1, \cdots, \sum_{k=0}^{d} n_k ,
\end{align}
and $\phi^{i_k}(t)$ receives $(k+1)$-th order derivative.  
Note that the numbering and the order of time derivative are off by $1$ for later convenience.  
We introduce the notation
\be \label{Qnot} Q^{i_0}_{00} \equiv \phi^{i_0} ,\quad
Q^{i_1}_{10} \equiv \phi^{i_1} ,\quad
\cdots \quad
Q^{i_d}_{d0} \equiv \phi^{i_d} , \ee 
and the auxiliary variables to rewrite the Lagrangian as
\begin{align} \label{Lageqi}
L_{\rm eq} &= L(
Q^{i_0}_{00}, \dot Q^{i_0}_{00}; 
Q^{i_1}_{10}, Q^{i_1}_{11}, \dot Q^{i_1}_{11}; 
Q^{i_2}_{20}, Q^{i_2}_{21}, Q^{i_2}_{22}, \dot Q^{i_2}_{22}; 
\cdots ;
Q^{i_d}_{d0}, Q^{i_d}_{d1}, \cdots , Q^{i_d}_{dd}, \dot Q^{i_d}_{dd}) \notag\\
&~~~~+ \lambda^{i_1}_{10} (\dot Q^{i_1}_{10} - Q^{i_1}_{11}) \notag\\
&~~~~+ \lambda^{i_2}_{20} (\dot Q^{i_2}_{20} - Q^{i_2}_{21}) + \lambda^{i_2}_{21} (\dot Q^{i_2}_{21} - Q^{i_2}_{22}) \notag\\
&~~~~+ \cdots \notag\\
&~~~~+ \lambda^{i_d}_{d0} (\dot Q^{i_d}_{d0} - Q^{i_d}_{d1}) + \lambda^{i_d}_{d1} (\dot Q^{i_d}_{d1} - Q^{i_d}_{d2}) + \cdots + \lambda^{i_d}_{d,d-1} (\dot Q^{i_d}_{d,d-1} - Q^{i_d}_{dd}) .
\end{align}
Therefore, we have $\{ Q, \lambda \}$ and their canonical momenta $\{ P, \rho \}$ which we classify as
\begin{align} 
q^i &\equiv (Q^{i_0}_{00}), & 
p_i &\equiv (P^{i_0}_{00}) ,\notag\\
\tilde Q^{(0)}_{I_1} &\equiv (Q^{i_1}_{11}, Q^{i_2}_{22}, \cdots, Q^{i_d}_{dd}) ,&
\tilde P^{(0)}_{I_1} &\equiv (P^{i_1}_{11}, P^{i_2}_{22}, \cdots, P^{i_d}_{dd}) ,\notag\\
\Q&\equiv
\bem
Q^{i_1}_{10} \\
Q^{i_2}_{20} & Q^{i_2}_{21} \\ 
\vdots & & \ddots \\
Q^{i_d}_{d0} & Q^{i_d}_{d1} & \cdots & Q^{i_d}_{d,d-1}  
\eem
,& 
\P&\equiv
\bem
P^{i_1}_{10} \\
P^{i_2}_{20} & P^{i_2}_{21} \\ 
\vdots & & \ddots \\
P^{i_d}_{d0} & P^{i_d}_{d1} & \cdots & P^{i_d}_{d,d-1}  
\eem
,\notag\\
\L&\equiv 
\bem
\lambda^{i_1}_{10} \\ 
\lambda^{i_2}_{20} & \lambda^{i_2}_{21} \\
\vdots & & \ddots \\
\lambda^{i_d}_{d0} & \lambda^{i_d}_{d1} & \cdots & \lambda^{i_d}_{d,d-1} 
\eem
,&
\R&\equiv
\bem
\rho^{i_1}_{10} \\ 
\rho^{i_2}_{20} & \rho^{i_2}_{21} \\
\vdots & & \ddots \\
\rho^{i_d}_{d0} & \rho^{i_d}_{d1} & \cdots & \rho^{i_d}_{d,d-1} 
\eem
,
\end{align}
where $I_1=(i_1,i_2,\cdots,i_d)$.
The total number of the canonical variables is thus a priori 
\begin{align} \label{Ncan}
N_{\rm can} &= 2 \sum_{k=0}^{d} (k+1) n_k + 2 \sum_{k=1}^{d} k n_k \notag\\
&= 2 \sum_{k=0}^{d} n_k + 4 \sum_{k=1}^{d} kn_k . 
\end{align}
Below we consider how to remove $4 \sum_{k=1}^{d} kn_k$ by constraints.

\subsection{Hamiltonian analysis}
\label{ssec:Ham3}

The canonical momenta are defined as 
\be p_i=L_i, \quad
\tilde P^{(0)}_{I_1}=L_{I_1}, \quad
\P = \L,\quad
\R = \0 , \ee
where 
$L_{I_1} \equiv \pa L / \pa \dot {\tilde Q}^{(0)}_{I_1}$.
First, from the latter two equations we obtain the primary constraints 
\be \label{conbPhibPhibar} \bPhi \equiv \P - \L \approx \0 , \quad 
\bar\bPhi \equiv \R \approx \0. \ee
As we shall see, they are second class constraints and thus constrain only $\L$ and $\R$.
Next we focus on the former two equations.
The $q^i$ and $Q^I$ sectors are parallel to those in the previous section.
Thus we assume $\det L_{ij} \neq 0$, and impose the first degeneracy condition 
\be \label{deg1i} L_{I_1J_1} - L_{I_1i} L^{ij} L_{jJ_1} = 0 , \ee
which is equivalent to the additional primary constraints
\be \label{Psii} \tilde\Psi^{(0)}_{I_1} \equiv \tilde P^{(0)}_{I_1} - \tilde F^{(0)}_{I_1}(p_i,x) \approx 0 , \ee
where
\be \tilde\Psi^{(0)}_{I_1} = (\Psi^{i_1}_{11}, \Psi^{i_2}_{22}, \cdots, \Psi^{i_d}_{dd}), \ee
and $x = (q^i,\tilde{Q}^{(0)}_{I_1},\Q)$.

To write down the total Hamiltonian in a simpler form we introduce the notation in addition to $\tilde Q^{(0)}_{I_1}\equiv (Q^{i_1}_{11}, Q^{i_2}_{22}, \cdots, Q^{i_d}_{dd})$ 
\begin{align}  
\tilde Q^{(1)}_{I_1} &\equiv (Q^{i_1}_{10}, Q^{i_2}_{21}, Q^{i_3}_{32}, \cdots, Q^{i_d}_{d,d-1}) , \notag\\
\tilde Q^{(2)}_{I_2} &\equiv (Q^{i_2}_{20}, Q^{i_3}_{31}, \cdots, Q^{i_d}_{d,d-2}) ,\notag\\
&\hspace{2mm}\vdots \notag\\
\tilde Q^{(d-1)}_{I_{d-1}} &\equiv (Q^{i_{d-1}}_{d-1,0}, Q^{i_d}_{d1}) , \notag\\
\tilde Q^{(d)}_{I_d} &\equiv (Q^{i_d}_{d0}) ,
\end{align}
which decompose the matrix $\Q$ into $d$ vectors, picking up the arguments from left top to right down.  
Here $I_k=(i_k,\cdots,i_d)$, and thus we can decompose  
\be \tilde Q^{(k)}_{I_k} = (Q^{i_k}_{k0}, \tilde Q^{(k)}_{I_{k+1}}) ,\quad
\tilde Q^{(k)}_{I_{k+1}} =(Q^{i_{k+1}}_{k+1,1},Q^{i_{k+2}}_{k+2,2},\cdots,Q^{i_{d}}_{d,d-k}), \ee
which we exploit below to isolate the first argument. 
We also define $\tilde P^{(k)}_{I_k}$ in the same way:
\be \tilde P^{(k)}_{I_k} \equiv (P^{i_k}_{k0},P^{i_{k+1}}_{k+1,1}, \cdots, P^{i_d}_{d,d-k} ) . \ee

With this notation, the Lagrangian \eqref{Lageqi} simplifies as
\be \label{Lageqit} L_{\rm eq} = L( q^i, \dot q^i ; \tilde Q^{(0)}_{I_1}, \dot{\tilde Q}^{(0)}_{I_1} ; \tilde Q^{(1)}_{I_1}, \tilde Q^{(2)}_{I_2}, \cdots , \tilde Q^{(d)}_{I_d} ) 
+ \sum_{k=1}^d \tilde \lambda^{(k)}_{I_k} ( \dot{\tilde Q}^{(k)}_{I_k} - \tilde Q^{(k-1)}_{I_k} ) . \ee
The total Hamiltonian is then given by
\begin{align} \label{HTarb}
H_T &= H + \mu_\alpha \Phi_\alpha + \bar\mu_\alpha \bar\Phi_\alpha + \tilde\nu^{(0)}_{I_1}\tilde\Psi^{(0)}_{I_1}, \notag\\ 
H &= H_0 + \tilde Q^{(0)}_{I_1} \tilde P^{(1)}_{I_1} + \tilde
 Q^{(1)}_{I_2} \tilde P^{(2)}_{I_2} + \cdots + \tilde Q^{(d-1)}_{I_d}
 \tilde P^{(d)}_{I_d} , \notag\\
H_0 &= \dot{\tilde Q}^{(0)}_{I_1} \tilde P^{(0)}_{I_1} + \dot q^ip_i - L , 
\end{align}
where $\Phi_\alpha=(\Phi^{i_1}_{10},\Phi^{i_2}_{20},\Phi^{i_2}_{21},\cdots,\Phi^{i_d}_{d0},\cdots,\Phi^{i_d}_{d,d-1})$ denotes the $\sum_{k=1}^{d} k n_k$ constraints, and so does $\bar\Phi_\alpha$.
Clearly, the linear terms $\tilde Q^{(k-1)}_{I_k} \tilde P^{(k)}_{I_k}$ cause Ostrogradsky instabilities.
Below we show how to remove them by imposing constraints to $\tilde P^{(k)}_{I_k}$.

The consistency conditions $\dot{\bar\Phi}_\alpha\approx 0$ and $\dot\Phi_\alpha\approx 0$ respectively give 
\be \mu_\alpha \approx 0, \quad 
\bar\mu_\alpha \approx \{\Phi_\alpha, H \} + \tilde\nu^{(0)}_{I_1} \{ \Phi_\alpha, \tilde\Psi^{(0)}_{I_1} \} , \ee
which determine $\mu_\alpha$ and $\bar\mu_\alpha$ once $\tilde\nu^{(0)}_{I_1}$ are fixed. 
Since the consistency condition for $\tilde \Psi^{(0)}_{I_1}$ is given by 
\be 0 \approx \dot{\tilde\Psi}^{(0)}_{I_1} = \{\tilde \Psi^{(0)}_{I_1},H\} + \tilde\nu^{(0)}_{J_1} \{ \tilde \Psi^{(0)}_{I_1}, \tilde \Psi^{(0)}_{J_1} \} , \ee
we impose the second degeneracy condition as 
\be \label{deg2d} \{ \tilde \Psi^{(0)}_{I_1}, \tilde \Psi^{(0)}_{J_1} \} = 0 , \ee
and we obtain secondary constraints
\be \label{con2d} \tilde \Upsilon^{(1)}_{I_1} \equiv - \{ \tilde \Psi^{(0)}_{I_1} ,H\} 
= \tilde P^{(1)}_{I_1} - G^{(1)}_{I_1}(p^i,x) \approx 0 , \ee
where 
\be \tilde \Upsilon^{(1)}_{I_1} 
\equiv (\Upsilon^{i_1}_{10},\Upsilon^{i_2}_{21}, \cdots, \Upsilon^{i_d}_{d,d-1} ) . \ee

Recalling that this notation allows us to isolate the first argument as $\tilde \Upsilon^{(1)}_{I_1} = (\Upsilon^{i_1}_{10}, \tilde \Upsilon^{(1)}_{I_2})$,
the consistency condition for $\tilde \Upsilon^{(1)}_{I_1} \approx 0$ is given by
\begin{align} \label{cncn1}
0 &\approx \dot\Upsilon^{i_1}_{10} = \{ \Upsilon^{i_1}_{10},H \} + \tilde\nu^{(0)}_{J_1} \{ \Upsilon^{i_1}_{10}, \tilde\Psi^{(0)}_{J_1} \} , \notag\\
0 &\approx \dot{\tilde\Upsilon}^{(1)}_{I_2} = \{\tilde\Upsilon^{(1)}_{I_2},H\} + \tilde\nu^{(0)}_{J_1} \{ \tilde\Upsilon^{(1)}_{I_2}, \tilde\Psi^{(0)}_{J_1} \} .
\end{align}
Since $\Upsilon^{i_1}_{10}$ fixes $P^{i_1}_{10}$ or the lowest problematic momentum for $\phi^{i_1}$ sector, we would like to avoid generating further constraints from $\dot\Upsilon^{i_1}_{10} \approx 0$.  
In other words, we do not need further constraint as the Hamiltonian does not contain linear term such as $Q^{i_1}_{10}P$ with some momentum $P$.
Therefore the first equation of \eqref{cncn1} gives $n_1$ equations between $\tilde\nu^{(0)}_{J_1}$.
In contrast, we would like to have further constraints from $\dot{\tilde\Upsilon}^{(1)}_{I_2}\approx 0$ to eliminate remaining linear terms coming from $\phi^{i_k}$ sectors with $k\geq 2$. 
We thus impose the third degeneracy condition
\be \label{deg3d} \{ \tilde\Upsilon^{(1)}_{I_2}, \tilde\Psi^{(0)}_{J_1} \} = 0. \ee
As we have discussed in \eqref{deg3I},
this is not the most general condition for \eqref{cncn1} to determine only $n_1$ component of $\tilde\nu^{(0)}_{I_1}$.
Analysis in more general case is definitely interesting, but becomes highly complicated and
is beyond the scope of this paper.
Thus, we impose~\eqref{deg3d}. 
Then, the second equation of \eqref{cncn1} yields the tertiary constraints
\be \label{con3d} \tilde\Upsilon^{(2)}_{I_2} \equiv - \{\tilde\Upsilon^{(1)}_{I_2}, H\} 
= \tilde P^{(2)}_{I_2} - \tilde G^{(2)}_{I_2}(p^i,x) \approx 0 .  \ee

By induction, for the constraints 
\be \label{conk} \tilde\Upsilon^{(k)}_{I_k} \equiv - \{\tilde\Upsilon^{(k-1)}_{I_k}, H\} 
= \tilde P^{(k)}_{I_k} - \tilde G^{(k)}_{I_k}(p^i,x) \approx 0 ,\ee 
we decompose the consistency conditions as
\begin{align} \label{cncn2}
0 &\approx \dot\Upsilon^{i_k}_{k0} = \{ \Upsilon^{i_k}_{k0},H \} + \tilde\nu^{(0)}_{J_1} \{ \Upsilon^{i_k}_{k0}, \tilde\Psi^{(0)}_{J_1} \} , \notag\\
0 &\approx \dot{\tilde\Upsilon}^{(k)}_{I_{k+1}} = \{\tilde\Upsilon^{(k)}_{I_{k+1}},H\} + \tilde\nu^{(0)}_{J_1} \{ \tilde\Upsilon^{(k)}_{I_{k+1}}, \tilde\Psi^{(0)}_{J_1} \} ,
\end{align}
and impose the degeneracy conditions
\be \label{degkd}
\{ \tilde\Upsilon^{(k)}_{I_{k+1}}, \tilde\Psi^{(0)}_{J_1} \} = 0 , \ee
to obtain the constraints 
\be \label{conkd} \tilde\Upsilon^{(k+1)}_{I_{k+1}} \equiv - \{\tilde\Upsilon^{(k)}_{I_{k+1}}, H\} 
= \tilde P^{(k+1)}_{I_{k+1}} - \tilde G^{(k+1)}_{I_{k+1}}(p^i,x) \approx 0 , \ee
for $k=2,\cdots,d-1$.

The constraints \eqref{con2d}, \eqref{con3d}, \eqref{conk}, \eqref{conkd} form a matrix
\be \label{bUpsilon} \bUp \equiv \P - \vG \approx 0 \ee
where
\be
\bUp\equiv
\bem
\Upsilon^{i_1}_{10} \\
\Upsilon^{i_2}_{20} & \Upsilon^{i_2}_{21} \\ 
\vdots & & \ddots \\
\Upsilon^{i_d}_{d0} & \Upsilon^{i_d}_{d1} & \cdots & \Upsilon^{i_d}_{d,d-1}  
\eem
,\quad 
\vG\equiv
\bem
G^{i_1}_{10} \\
G^{i_2}_{20} & G^{i_2}_{21} \\ 
\vdots & & \ddots \\
G^{i_d}_{d0} & G^{i_d}_{d1} & \cdots & G^{i_d}_{d,d-1}  
\eem
.
\ee
We then arrive at the consistency condition for the last constraint $\tilde\Upsilon^{(d)}_{I_d}=\Upsilon^{i_d}_{d0}$ 
\be \label{cncnd} 0 \approx \dot \Upsilon^{i_d}_{d0} = \{ \Upsilon^{i_d}_{d0},H \} + \tilde\nu^{(0)}_{J_1} \{ \Upsilon^{i_d}_{d0}, \tilde\Psi^{(0)}_{J_1} \} . \ee

After the above procedure, the constraints~\eqref{bUpsilon} 
fix the linear momentum terms in the Hamiltonian \eqref{HTarb}, and we are left with the consistency conditions
\be \label{Upcncn} \{ \Upsilon^{i_k}_{k0},H \} + \tilde\nu^{(0)}_{I_1} \{ \Upsilon^{i_k}_{k0}, \tilde\Psi^{(0)}_{I_1} \} \approx 0, \quad (k=1,\cdots, d). \ee
Therefore, if this set of equations determine the Lagrange multipliers $\tilde\nu^{(0)}_{I_1}$, we complete the Dirac algorithm.

In parallel to \eqref{phaseI}, we can list all the phase space variables of the current system~\eqref{Lageqi} as
\be \label{phasei}
\begin{array}{cccc}
  \tilde Q^{(0)}_{I_1} & q^i & \Q & \boxed{\L}   \\ 
  \boxed{\tilde P^{(0)}_{I_1}} & p_i & \boxed{\P} & \boxed{\R}  
\end{array} ,
\ee
where the boxed variables are fixed in terms of other variables via constraints obtained so far.
Nevertheless, as a natural generalization of the results obtained in \S
\ref{sec:multi}, we are interested in the case where the number of
degrees of freedom matches the number of variables by removing all the
ghosts associated with the 
canonical variables which correspond to the higher-than-first time
derivatives of the original variables, and all the constraints are second class.
Such canonical variables come from $\tilde Q^{(0)}_{I_1} \equiv (Q^{i_1}_{11}, Q^{i_2}_{22}, \cdots, Q^{i_d}_{dd})$ and $\Q$.
We can combine $\tilde Q^{(0)}_{I_1}$ and $\Q$, 
and list up them as a larger matrix
\be 
\bem
Q^{i_1}_{10} & Q^{i_1}_{11} \\
Q^{i_2}_{20} & Q^{i_2}_{21} & Q^{i_2}_{22} \\ 
Q^{i_3}_{30} & Q^{i_3}_{31} & Q^{i_3}_{32} & Q^{i_3}_{33} \\ 
\vdots &  & & & \ddots \\
Q^{i_d}_{d0} & Q^{i_d}_{d1} & Q^{i_d}_{d2} & \cdots & Q^{i_d}_{d,d-1} & Q^{i_d}_{dd} 
\eem.
\ee
The first two columns are the original variables and their first-order time derivatives, and the remaining part
\be \label{Qprime} \Q' \equiv
\bem
0 \\
Q^{i_2}_{22} & 0 \\ 
Q^{i_3}_{32} & Q^{i_3}_{33} & 0\\
\vdots & & \ddots & \ddots \\
Q^{i_d}_{d2} & Q^{i_d}_{d3} & \cdots & Q^{i_d}_{dd} & 0
\eem ,
\ee
is the variables that we would like to fix by invoking additional constraints.
Here, we keep a row and a column of zeros in the definition of $\Q'$ 
and make its dimension as the same as the other matrices denoted by the bold font.

We thus require an additional degeneracy condition 
\be \label{dega2} \{ \Upsilon^{i_k}_{k0}, \tilde\Psi^{(0)}_{I_1} \} = 0 , \quad (k=2,\cdots, d), \ee
with which \eqref{Upcncn} yields additional constraints 
\be \Omega^{i_k}_{k0} \equiv - \{ \Upsilon^{i_k}_{k0},H \} \approx 0 , \quad (k=2,\cdots, d). \ee
Note that, analogous to \S\ref{sec:multi}, we do not impose the degeneracy condition for $i_1$ component. 
The number of the constraints is $n_2+\cdots n_d$, which is the same as the number of the nonvanishing components of the first column of $\Q'$ in \eqref{Qprime}.
The consistency conditions for $\Omega^{i_k}_{k0}$ are given by
\be \{ \Omega^{i_k}_{k0},H \} + \tilde\nu^{(0)}_{I_1} \{ \Omega^{i_k}_{k0}, \tilde\Psi^{(0)}_{I_1} \} \approx 0, \quad (k=2,\cdots, d) . \ee

To obtain a sufficient number of constraints, we further impose degeneracy conditions for $k=3,\cdots, d$
\be \label{dega3} \{ \Omega^{i_k}_{k0}, \tilde\Psi^{(0)}_{I_1} \} = 0 ,\quad (k=3,\cdots, d) ,\ee
and we obtain constraints 
\be \Omega^{i_k}_{k1}\equiv -\{ \Omega^{i_k}_{k0},H \} \approx 0 ,\quad (k=3,\cdots, d), \ee
whose consistency conditions are given by
\be \{ \Omega^{i_k}_{k1},H \} + \tilde\nu^{(0)}_{I_1} \{ \Omega^{i_k}_{k1}, \tilde\Psi^{(0)}_{I_1} \} \approx 0, \quad (k=3,\cdots, d) . \ee

We continue the process $k-2$ times and impose 
\be \label{degal} \{ \Omega^{i_\ell}_{\ell k}, \tilde\Psi^{(0)}_{I_1} \} = 0 ,\quad (k=3,\cdots,d-3;~ \ell=k+1,\cdots, d) ,\ee 
until we obtain a set of constraints
\be \label{bOmega}
\bOm\equiv
\bem
0 \\
\Omega^{i_2}_{20} & 0 \\
\Omega^{i_3}_{30} & \Omega^{i_3}_{31} & 0 \\ 
\vdots & & \ddots & \ddots \\
\Omega^{i_d}_{d0} & \Omega^{i_d}_{d1} & \cdots & \Omega^{i_d}_{d,d-2} & 0
\eem ,
\ee
which precisely corresponds to $\Q'$ in \eqref{Qprime}.  
The remaining consistency conditions corresponding to the diagonal zero components of \eqref{bOmega} are given by
\be \label{hatOm} \{ \hat\Omega^{(1)}_{I_1},H \} + \tilde\nu^{(0)}_{I_1} \{ \hat\Omega^{(1)}_{I_1}, \tilde\Psi^{(0)}_{I_1} \} \approx 0 , \ee
where we defined 
\be \hat\Omega^{(1)}_{I_1} \equiv (\Upsilon^{i_1}_{10},\tilde\Omega^{(2)}_{I_2}) , \ee
and
\be \tilde\Omega^{(2)}_{I_2} \equiv (\Omega^{i_2}_{20}, \Omega^{i_3}_{31}, \cdots, \Omega^{i_d}_{d,d-2}). \ee

All the degeneracy conditions we imposed above are 
\eqref{deg1i}, \eqref{deg2d}, \eqref{deg3d}, \eqref{degkd}, \eqref{dega2}, \eqref{dega3}, \eqref{degal}, 
which are
summarized as  
\begin{align} \label{degUpOm}
L_{I_1J_1} - L_{I_1i} L^{ij} L_{jJ_1} &= 0 , \notag\\
\{ \tilde \Psi^{(0)}_{I_1}, \tilde \Psi^{(0)}_{J_1} \} &= 0 ,\notag\\
\{ \bUp, \tilde\Psi^{(0)}_{J_1} \} &= 0 \quad \text{except} \quad \Upsilon^{i_1}_{10} , \notag\\
\{ \bOm, \tilde\Psi^{(0)}_{J_1} \} &= 0 \quad \text{except} \quad \tilde\Omega^{(2)}_{I_2} .
\end{align} 
Now we require \eqref{hatOm} determines all the Lagrange multipliers $\tilde\nu^{(0)}_{I_1}$, and complete the Dirac algorithm.
As a generalization of \eqref{lastmulti}, we define a matrix
\be Z_{I_1 J_1} \equiv \{\hat\Omega^{(1)}_{I_1}, \tilde\Psi^{(0)}_{J_1} \}. \ee
One can show that $Z_{i_k, j_\ell}=0$ for $k < \ell$ [see also \eqref{ABP} below].
Thus, the necessary and sufficient condition for
\eqref{hatOm} to determine all $\tilde\nu^{(0)}_{I_1}$ is that each $(i_k,j_k)$ submatrix is nondegenerate
\begin{align} \label{lastarb}
\det Z_{i_k j_k} &\neq 0.
\end{align}
With the degeneracy conditions \eqref{degUpOm},
we obtain the constraints $(\bPhi, \bar\bPhi, \tilde \Psi^{(0)}_{I_1}, \bUp, \bOm)$
which are given in \eqref{conbPhibPhibar}, \eqref{Psii}, \eqref{bUpsilon}, \eqref{bOmega}.
The correspondence between the canonical variables and the constraints that fix them are
\be
\L: \bPhi, \quad
\R: \bar\bPhi, \quad
\tilde P^{(0)}_{I_1} : \tilde \Psi^{(0)}_{I_1} ,\quad
\P: \bUp , \quad
\Q': \bOm.
\ee
While we do not show that the constraints $\bOm$ fixes the variables $\Q'$ explicitly, the correspondence is reasonable as the $\Q'$ amounts to higher derivatives and the number of constraints and variables precisely match.
The number of constraints are respectively
\begin{align} \label{Ncon}
\bPhi : \sum_{k=1}^d kn_k ,  \quad
\bar\bPhi : \sum_{k=1}^d kn_k , \quad
\tilde \Psi^{(0)}_{I_1} : \sum_{k=1}^d n_k , \quad
\bUp : \sum_{k=1}^d kn_k  , \quad
\bOm : \sum_{k=1}^d (k-1)n_k .
\end{align}
To count the number of degrees of freedom we shall classify them into first class and second class constraints.

While the correspondence to canonical variables is transparent for the combination $(\bPhi, \bar\bPhi, \tilde \Psi^{(0)}_{I_1}, \bUp, \bOm)$, it is not the best combination for counting the number of degrees of freedom as the Dirac matrix is not simple. 
Let us focus on $(\tilde \Psi^{(0)}_{I_1}, \bUp, \bOm)$ and consider a more useful basis.  
These constraints are connected each other by chains of Poisson brackets.  
Let us list them as
\be \label{Tri1}
\begin{array}{cccccc|cccccc}
 & & & & & \Psi^{i_1}_{11} & \Upsilon^{i_1}_{10} \\
 & & & & \Psi^{i_2}_{22} & \Upsilon^{i_2}_{21} & \Upsilon^{i_2}_{20} & \Omega^{i_2}_{20} \\
 & & & \Psi^{i_3}_{33} & \Upsilon^{i_3}_{32} & \Upsilon^{i_3}_{31} & \Upsilon^{i_3}_{30} & \Omega^{i_3}_{30} & \Omega^{i_3}_{31} \\
 & & \Psi^{i_4}_{44} & \Upsilon^{i_4}_{43} & \Upsilon^{i_4}_{42} & \Upsilon^{i_4}_{41} & \Upsilon^{i_4}_{40} & \Omega^{i_4}_{40} & \Omega^{i_4}_{41} & \Omega^{i_4}_{42} \\ 
 & \rddots & \rddots & & & \vdots & \vdots & \vdots & & & \ddots \\
\Psi^{i_d}_{dd} & \Upsilon^{i_d}_{d,d-1} & & \cdots & & \Upsilon^{i_d}_{d1} & \Upsilon^{i_d}_{d0} & \Omega^{i_d}_{d0} & & \cdots & & \Omega^{i_d}_{d,d-2} \\
\end{array}
\ee
Each row is connected by a chain of Poisson brackets.
Starting from the most left component of $\tilde \Psi^{(0)}_{I_1}$, the next right component is defined by taking a Poisson bracket with $-H$, and we continue to proceed to the right component until we arrive at the most right component of $\hat\Omega^{(1)}_{I_1}$ ending with nonvanishing Poisson bracket with corresponding component of $\tilde \Psi^{(0)}_{I_1}$.
Taking the bottom row of \eqref{Tri1} as an example, we have 
$\Upsilon^{i_d}_{d,d-1} = -\{\Psi^{i_d}_{dd}, H\}$,
$\Upsilon^{i_d}_{d,d-2} = -\{\Upsilon^{i_d}_{d,d-1}, H\}$, $\cdots$,
$\Omega^{i_d}_{d,d-2} = -\{\Omega^{i_d}_{d,d-3}, H\}$, and 
$\det \{\Omega^{i_d}_{d,d-2}, \Psi^{i_d}_{dd} \} \neq 0$.
To make use of the structure of Poisson brackets, it is more useful to divide the constraints by the vertical line shown in \eqref{Tri1} rather than distinguishing them by $\Upsilon, \Omega$ notation.
We thus reclassify and relabel them as
\be
\begin{array}{cccccc|cccccc}
 & & & & & \Psi^{i_1}_{11} & \Upsilon^{i_1}_{10} \\
 & & & & \Psi^{i_2}_{22} & A^{i_2}_{1} & B^{i_2}_{1} & \Omega^{i_2}_{20} \\
 & & & \Psi^{i_3}_{33} & A^{i_3}_{1} & A^{i_3}_{2} & B^{i_3}_{2} & B^{i_3}_{1} & \Omega^{i_3}_{31} \\
 & & \Psi^{i_4}_{44} & A^{i_4}_{1} & A^{i_4}_{2} & A^{i_4}_{3} & B^{i_4}_{3} & B^{i_4}_{2} & B^{i_4}_{1} & \Omega^{i_4}_{42} \\ 
 & \rddots & \rddots & & & \vdots & \vdots & \vdots & & \ddots & \ddots \\
\Psi^{i_d}_{dd} & A^{i_d}_{1} & & \cdots & & A^{i_d}_{d-1} & B^{i_d}_{d-1} & B^{i_d}_{d-2} & \cdots & & B^{i_d}_{1} & \Omega^{i_d}_{d,d-2} \\
\end{array}
\ee
The chains of Poisson brackets are then rewritten as
\begin{align}
\Upsilon^{i_1}_{10} &= -\{ \Psi^{i_1}_{11} , H \} ,\notag\\
A^{i_k}_{1} &= - \{ \Psi^{i_k}_{kk} , H \} , \qquad~\, (k=2,\cdots, d), \notag\\
A^{i_k}_{a+1} &= - \{ A^{i_k}_{a}, H \} , \qquad~~ (k=3,\cdots, d;~ a=1,\cdots, k-2), \notag\\
B^{i_k}_{k-1} &= -\{ A^{i_k}_{k-1}, H\} , \qquad (k=2,\cdots, d), \notag\\
B^{i_k}_{a} &= - \{ B^{i_k}_{a+1}, H \} , \qquad (k=3,\cdots, d;~ a=1,\cdots, k-2), \notag\\
\Omega^{i_k}_{k,k-2} &= -\{ B^{i_k}_{1}, H \}, \qquad~~ (k=2,\cdots, d).
\end{align}
Note that the last two degeneracy conditions of \eqref{degUpOm} read  
\be \{ A^{i_k}_a, \tilde\Psi^{(0)}_{I_1} \} = \{ B^{i_k}_a, \tilde\Psi^{(0)}_{I_1} \} = 0, \quad (a=1,\cdots, k-1). \ee
From these relations and the Jacobi identity, one can show that 
\begin{align} \label{ABP}
\{ A^{i_k}_a, B^{j_k}_a \} &= (-1)^{a+1} Z_{j_ki_k}, \notag\\ 
\{ A^{i_k}_a, B^{j_\ell}_b \} &= 0 , \quad (a<b  ~~\text{or}~~ 2\ell-b < 2 k-a) , \notag\\
\{ A^{i_k}_a, A^{j_\ell}_b \} &= 0 .
\end{align}

With the above basis, the Dirac matrix is given by
\be
\begin{array}{l|ccccccccc}
             & \bPhi & \bar\bPhi & \tilde\Psi & \hat\Omega & A^{j_2}_{1} & B^{j_2}_1 & \cdots & A^{j_d}_{d-1} & B^{j_d}_{d-1} \\ \hline
  \bPhi      & 0 & -\1 & * & * & * & * & * & * & *    \\
  \bar\bPhi  & \1 & 0 & 0 & 0 & 0 & 0 & 0 & 0 & 0   \\
  \tilde\Psi & * & 0 & 0 & -Z^T & 0 & 0 & 0 & 0 & 0  \\
  \hat\Omega & * & 0 & Z & * & * & * & * & * & *   \\
  A^{i_2}_1  & * & 0 & 0 & * & 0 & Z_{j_2i_2} & 0 & 0 & 0\\
  B^{i_2}_1  & * & 0 & 0 & * & -Z_{i_2j_2} & * & * & * & *  \\
  \vdots     & * & 0 & 0 & * & 0 & * & \ddots & \vdots & \vdots \\
  A^{i_d}_{d-1}  & * & 0 & 0 & * & 0 & * & \cdots & 0 & (-1)^d Z_{j_di_d}  \\
  B^{i_d}_{d-1}  & * & 0 & 0 & * & 0 & * & \cdots & (-1)^{d+1} Z_{i_dj_d}  & * \\ 
\end{array} 
\ee
The determinant of the Dirac matrix is thus given by
\be (\det Z_{i_1 j_1})^2 \prod_{k=2}^d (\det Z_{i_k j_k})^4 , \ee
which is nonvanishing by virtue of \eqref{lastarb}.  
Hence, all the constraints are second class, whose total number is given by summing up \eqref{Ncon}
\be N_{\rm 2nd} = 4 \sum_{k=0}^d k n_k .  \ee
Using \eqref{Ncan}, the number of degrees of freedom is 
\be N_{\rm DOF} = \f{1}{2} ( N_{\rm can} - N_{\rm 2nd} ) = \sum_{k=0}^d n_k . \ee

\subsection{Euler-Lagrange equation}
\label{ssec:ELeq3}

The Euler-Lagrange equation for the Lagrangian \eqref{Lageqit} can be written as
\begin{align}
\label{ELeq1i} \dot L_{i} - L_{q^i} &= 0 ,\\
\label{ELeq2i} \dot L_{I_1} - L_{\tilde Q^{(0)}_{I_1}} + \tilde\lambda^{(1)}_{I_1} &= 0 ,\\
\label{ELeq3i} L_{\tilde Q^{(k)}_{I_{k+1}}} - \tilde\lambda^{(k+1)}_{I_{k+1}} - \dot{\tilde\lambda}^{(k)}_{I_{k+1}} &= 0, \qquad (k=1, \cdots, d-1), \\
\label{ELeq4i} L_{Q^{i_k}_{k0}} - \dot\lambda^{i_k}_{k0} &= 0, \qquad (k=1,\cdots , d) , \\
\label{ELeq5i} \tilde Q^{(k)}_{I_{k+1}} - \dot{\tilde Q}^{(k+1)}_{I_{k+1}} &= 0, \qquad (k = 0, \cdots, d-1), \\
\label{ELeq6i} Q^{i_k}_{k1} - \dot Q^{i_k}_{k0} &= 0, \qquad (k=1,\cdots , d),
\end{align}
where we recall that $L_i\equiv L_{\dot q^i}$ and $L_{I_1}\equiv L_{\dot{\tilde Q}^{(0)}_{I_1}}$.
The above equations \eqref{ELeq1i}--\eqref{ELeq6i} have the same structure as the equations \eqref{ELeq1I}--\eqref{ELeq8I}, though the numbers of two sets of equations are different:
\eqref{ELeq1i}, \eqref{ELeq2i}, \eqref{ELeq3i}, \eqref{ELeq5i}
correspond to 
\eqref{ELeq1I}, \eqref{ELeq2I}, \eqref{ELeq3I}, \eqref{ELeq6I}, respectively,
whereas
\eqref{ELeq4i} corresponds to \eqref{ELeq4I} and \eqref{ELeq5I}, and 
\eqref{ELeq6i} corresponds to~\eqref{ELeq7I} and \eqref{ELeq8I}.
We shall show below that their reduction to a second-order system is a natural generalization of the analysis in \S\ref{ssec:ELeq2}.

Following \S\ref{ssec:ELeq2}, first we focus on $\tilde\lambda^{(1)}_{I_1}
$ in \eqref{ELeq2i}.
While a priori \eqref{ELeq2i} implies that $\tilde\lambda^{(1)}_{I_1}$ depends on $\ddot{\tilde Q}^{(0)}_{I_1}$, with the first degeneracy condition \eqref{deg1i} or the additional primary constraints \eqref{Psii}, we can show that \eqref{ELeq1i} and \eqref{ELeq2i} read
\begin{align}
\label{lami1} &~~~~ \tilde\lambda^{(1)}_{I_1} = L_{\tilde Q^{(0)}_{I_1}} - \tilde F^{(0)}_{I_1p_i} L_{q^i} - \dot x \tilde F^{(0)}_{I_1x} , \\
\label{EOMiq} \E_i &\equiv \ddot q^i + F_{I_1p_i} \ddot{\tilde Q}^{(0)}_{I_1} - L^{ij} (L_{q^j} - \dot x L_{jx}) = 0 ,
\end{align}
the former of which corresponds to the secondary constraints \eqref{con2d}.

Next we focus on \eqref{ELeq3i} and \eqref{ELeq4i} with $k=1$.
We take a time derivative of \eqref{lami1}, $I_2$ component of which gives $\tilde\lambda^{(2)}_{I_2}$ from \eqref{ELeq3i} with $k=1$, and $i_1$ component of which gives EOM for $Q^{i_1}_{10}$ from \eqref{ELeq4i} with $k=1$.
Again, while they a priori depend on $\ddot{\tilde Q}^{(0)}_{I_1}$, with the second degeneracy condition \eqref{deg2d}, \eqref{lami1} implies $\tilde\lambda^{(1)}_{I_1} =\tilde G^{(1)}_{I_1} (L_i,x)$ and $\ddot{\tilde Q}^{(0)}_{I_1}$ dependences identically vanish.
From \eqref{ELeq3i} with $k=1$ we obtain 
\be \label{lami2} \tilde\lambda^{(2)}_{I_2} = L_{\tilde Q^{(1)}_{I_2}} - \tilde G^{(1)}_{I_2p_i} L_{q^i} - \dot x \tilde G^{(1)}_{I_2x} , \ee
which corresponds to the tertiary constraints $\tilde\lambda^{(2)}_{I_2}=G^{(2)}_{I_2}(L_i,x)$ in \eqref{con3d}. 
Also, from \eqref{ELeq4i} with $k=1$ we obtain EOM for $Q^{i_1}_{10}$ as
\be \label{EOM1} 0 = \E^{i_1}_{10} \equiv L_{Q^{i_1}_{10}} - G^{i_1}_{10,p_i} L_{q^i} - \dot x G^{i_1}_{10,x} = - \{ \Upsilon^{i_1}_{10}, H \}, \ee
which, recalling the notation \eqref{Qnot}, is the EOM for $\phi^{i_1}$, and shows that the first term of the most right hand side of the first equation of \eqref{cncn1} vanishes.

Inductively, for $k=2,\cdots d-1$, by using a time derivative of $\tilde\lambda^{(k)}_{I_k} = \tilde G^{(k)}_{I_k} (L_i,x)$ and the degeneracy conditions~\eqref{deg3d} and \eqref{degkd}, we can reduce \eqref{ELeq3i} and \eqref{ELeq4i} and obtain 
\begin{align} 
\label{lamik} \tilde\lambda^{(k+1)}_{I_{k+1}} &= L_{\tilde Q^{(k)}_{I_{k+1}}} - \tilde G^{(k)}_{I_{k+1}p_i} L_{q^i} - \dot x \tilde G^{(k)}_{I_{k+1}x} 
\equiv \tilde G^{(k+1)}_{I_{k+1}} (L_i,x), \\
\label{EOMk} 0 &= \E^{i_k}_{k0} \equiv L_{Q^{i_k}_{k0}} - G^{i_k}_{k0,p_i} L_{q^i} - \dot x G^{i_k}_{k0,x} = - \{ \Upsilon^{i_k}_{k0}, H \} ,
\end{align}
the latter of which is the EOM for $\phi^{i_k}$ and related to \eqref{cncn2}.
Finally, plugging a time derivative of \eqref{lamik} into \eqref{ELeq4i} with $k=d$ and using the degeneracy condition~\eqref{degkd} with $k=d-1$, we obtain EOM for $Q^{i_d}_{d0}=\phi^{i_d}$ as
\be \label{EOMd} 0 = \E^{i_d}_{d0} \equiv L_{Q^{i_d}_{d0}} - G^{i_d}_{d0,p_i} L_{q^i} - \dot x G^{i_d}_{d0,x} = - \{ \Upsilon^{i_d}_{d0}, H \} , \ee
which is related to \eqref{cncnd}.

We thus obtain EOMs for $q^i, \phi^{i_1}, \cdots, \phi^{i_d}$ as \eqref{EOMiq}, \eqref{EOMk}, \eqref{EOMd}, but they still contain higher derivatives. 
We can construct a set of EOMs with derivatives up to second-order as follows.
By virtue of the degeneracy condition \eqref{dega2}, $\E^{i_k}_{k0}$ for $k=2,\cdots, d$ are functions of $(L_i,x)$ and thus a time derivative of EOMs $\E^{i_k}_{k1}\equiv \dot \E^{i_k}_{k0}$ for $k=2,\cdots, d$ does not contain $\ddot{\tilde Q}^{(0)}_{I_1}$. 
We continue this procedure with the degeneracy conditions~\eqref{dega3} and \eqref{degal} to obtain a set of EOMs 
\be \label{mEs} 0 = \mE \equiv \bem
\E^{i_1}_{10} \\
\E^{i_2}_{20} & \E^{i_2}_{21} \\ 
\vdots & & \ddots \\
\E^{i_d}_{d0} & \E^{i_d}_{d1} & \cdots & \E^{i_d}_{d,d-1}  
\eem . \ee
Generalizing the logic for \eqref{dQnsol}--\eqref{dQasol}, we expect that in general the condition \eqref{lastarb} guarantees that we can solve \eqref{mEs} and 
express
\begin{align} 
\label{dtQsol} \dot{\tilde Q}^{(0)}_{I_1} &= \tilde F^{(0)}_{I_1} (\dot q^i, q^i, Q^{i_k}_{k0}, Q^{i_k}_{k1}), \\
\label{Qpsol} \Q' &= \mF' (\dot q^i,  q^i, Q^{i_k}_{k0}, Q^{i_k}_{k1}) ,
\end{align}
where $\mF'$ is a matrix with nonvanishing arguments corresponding to $\Q'$ defined in \eqref{Qprime}.
These equations are a generalization of \eqref{dQnsol}--\eqref{dQasol}.
From these equations, $\dot{Q}^{i_1}_{11}=\ddot\phi^{i_1}$ and $Q^{i_k}_{k2}=\ddot\phi^{i_k}$ for $k=2,\cdots, d$ can be written down in terms of derivatives up to first order. 
Taking a time derivative of \eqref{dtQsol} and plugging \eqref{dtQsol} and \eqref{Qpsol} we obtain 
\be \label{ddtQsol} \ddot{\tilde Q}^{(0)}_{I_1} = \tilde F^{(0)}_{I_1} (\ddot q^i, \dot q^i, q^i, Q^{i_k}_{k0}, Q^{i_k}_{k1}) . \ee
By substituting \eqref{dtQsol}--\eqref{ddtQsol} to \eqref{EOMiq} we obtain EOM containing at most $\ddot q^i$. 
We thus obtain a system of $\sum_{k=0}^d n_k$ EOMs that contain at most second-order derivatives.

\section{Conclusions and discussion}
\label{sec:conc}

In this paper, we have clarified how to construct no-ghost theory for general Lagrangians for point particle system involving arbitrary higher-order time derivatives. 
The first no-ghost theory involving third-order derivative was the quadratic model studied in \cite{Motohashi:2017eya}.
In \S\ref{sec:exs}, we provided the specific no-ghost theory that involves arbitrary higher-order derivative.
Then, in \S\ref{sec:multi}, we have derived the conditions for general Lagrangian involving third-order derivatives to possess only healthy DOFs.
As shown in \cite{Motohashi:2017eya}, 
in sharp contrast to theories with up to the second-order time derivatives in the
Lagrangian, eliminating linear dependence of canonical momenta in the
Hamiltonian is not sufficient for those with higher-than-second-order
derivatives, and that canonical coordinates corresponding to the higher
time-derivatives also need to be removed appropriately.  
In \cite{Motohashi:2017eya}, this process was confirmed for the quadratic model, 
and in \S\ref{sec:multi} we confirmed it for any Lagrangian involving third-order derivatives.
We have also
shown that, as long as these conditions are satisfied, the
Euler-Lagrange equations can be reduced to a system of second-order
differential equations, which is consistent with the absence of ghost
DOFs.  Finally, in \S\ref{sec:multi-high} we have extended these
analyses to general theories involving arbitrary higher-order derivatives. 
The caveat is that
we have concentrated on the cases, in which all of
the constraints are second class. If some of them are first class, the
analyses would be much more complicated and case-by-case analysis would
be necessary though such analyses are indispensable for gauge theories.
Nevertheless, 
by introducing adequate gauge fixing terms, 
first class constraints turn into second class ones, 
to which the analysis in the present paper would apply. 
We leave this kind of analysis as future work.

While our analysis is confined to the analytic mechanics for a system of point particles as the first step, 
it clarifies the essence of the construction of degenerate theories, and 
it is quite robust as they apply to
any Lagrangian involving arbitrary higher-order derivatives.
Furthermore, the analysis for field theory can be reduced into the one for the analytic mechanics
by exploiting ADM decomposition with a choice of direction of time.
After that, the result of the present paper will guide us how to construct ghost-free field theories 
with arbitrary higher-order derivatives. Actually, the
extension of our analysis to field theories with arbitrary higher-order
derivatives is quite interesting, for example, scalar (and vector)
fields in the Minkowski background, scalar-tensor theories,
vector-tensor theories, scalar-vector-tensor (TeVeS) theories, and even
a theory with fermionic degrees of freedom. Especially, it is
challenging to find a healthy theory with higher-order derivative terms,
which cannot be transformed to a theory with only up to first order
derivatives by invertible transformation \cite{Takahashi:2017zgr}. 
We also leave all of these topics as future work.

\acknowledgments

This work was supported in part by JSPS KEKENHI Grant Numbers,
JP17H06359 (H.M.), JP18K13565 (H.M.),
JP15K17632 (T.S.), JP17H06359 (T.S.), JP15H05888 (T.S.\ \& M.Y.),
JP25287054 (M.Y.), JP18H04579 (M.Y.).

\appendix

\section{Lagrangian with single third-order derivative}
\label{sec:sing}

In this Appendix, we consider the special case of the Lagrangian considered in \S\ref{sec:multi} with $\N=1,\A=0$.
In this case some part of degeneracy conditions are identically satisfied.  
While it is obvious that the equation corresponding to the second degeneracy conditions~\eqref{deg2I} is identically satisfied as $\{\Psi,\Psi\}=0$, it is more subtle to see the another equation $\{\Lambda,\Psi\}=0$ corresponding to the fourth degeneracy condition~\eqref{deg4I} is identically satisfied. 
Below we provide the proof of this equation.

The consistency condition for the tertiary constraint corresponding to \eqref{consisLam} reads
\be \label{con7} 0 \approx \dot \Lambda = \{ \Lambda, H\} +\nu\{\Lambda,\Psi\} . \ee
To obtain the quaternary constraint, we need
\be \label{P75} \{\Lambda,\Psi\} = F_\psi - I_R + I_{q^i} F_{p_i} - I_{p_i} F_{q^i} = 0 , \ee
corresponding to the fourth degeneracy condition~\eqref{deg4I}.
Actually, we can show that \eqref{P75}
identically holds by using Jacobi identity repeatedly:
\begin{align} \label{LPbk}
\{\Lambda,\Psi\} &= \{ -I, P_Q-F \} -\{\pi,F\} \notag\\
&= \{ \{ G, H_0+\pi R + P_R Q \} , P_Q - F \} - \{ \{ P_R,H_0 \}, P_Q-F \} - \{\pi, F\} \notag\\
&= - \{ \{ H_0+\pi R + P_R Q , P_Q-F \}, G \} - \{ \{ P_Q-F , G \}, H_0+\pi R + P_R Q \} \notag\\
&~~~\, - \{ \{ P_R,H_0 \}, P_Q-F \} - \{\pi, F\} \notag\\
&= - \{ -G+P_R, G \} - \{ \{ P_R,F \} , H_0+\pi R + P_R Q \} - \{ \{ P_R,H_0 \}, P_Q-F \} - \{\pi, F\} \notag\\ 
&= \{ G, P_R \} - \{ \{ P_R,F \} , H_0+\pi R + P_R Q \} - \{ \{ P_R,H_0 \}, P_Q-F \} - \{\pi, F\} \notag\\ 
&= \{ \{P_Q,H_0\}, P_R \} -\{ \{F,H_0\}, P_R \} - \{ \{F,\pi\}R, P_R \} - \{ \{F,P_R\}Q, P_R \} \notag\\
&~~~\, - \{ \{ P_R,F \} , H_0 \} - \{ \{ P_R,F \} , \pi R \} - \{ \{ P_R,F \} , P_R Q \} \notag\\
&~~~\, - \{ \{ P_R,H_0 \}, P_Q \} + \{ \{ P_R,H_0 \}, F \} - \{\pi, F\} \notag\\ 
&= - \{ F,\pi \} - \{ \{ F,\pi \} , P_R \}R - \{ \{ P_R,F \} , \pi \}R - \{ \pi,F \} \notag\\
&= 0 .
\end{align}
We thus have the quaternary constraint  
\begin{align} \label{Phi8} 
0\approx \Omega &\equiv - \{ \Lambda, H \} = - L_\psi + I_{p_i} L_{q^i} + \dot x I_x , 
\end{align}
corresponding to \eqref{con4I}.

\bibliographystyle{JHEPmod}
\bibliography{ref-3deriv}

\end{document}